\documentclass{emulateapj}
\usepackage{graphicx}
\usepackage[space]{grffile}
\usepackage{latexsym}
\usepackage{textcomp}
\usepackage{longtable}
\usepackage{multirow,booktabs}
\usepackage{amsfonts,amsmath,amssymb}
\usepackage{natbib}
\usepackage{url}
\usepackage{hyperref}
\hypersetup{colorlinks=false,pdfborder={0 0 0}}
\newif\iflatexml\latexmlfalse
\usepackage[utf8]{inputenc}
\usepackage[english]{babel}

\newcommand{\msun}{\,\mathrm{M}_\odot} 
\newcommand{\mjup}{\,\mathrm{M}_{\rm Jup}}
\newcommand{\yr}{\,\mathrm{yr}} 
\newcommand{\pc}{\,\mathrm{pc}}
\newcommand{\AU}{\,\mathrm{AU}} 

\shorttitle{Dynamics of tidally captured planets in the Galactic Center}
\shortauthors{Trani et al.}

\begin{document}


\title{Dynamics of tidally captured planets in the Galactic Center}


\author{Alessandro A. Trani\altaffilmark{1,2,$\star$}}

\author{Michela Mapelli\altaffilmark{2}}

\author{Mario Spera\altaffilmark{2}}

\author{Alessandro Bressan\altaffilmark{1,2}}

\altaffiltext{1}{Scuola Internazionale Superiore di Studi Avanzati (SISSA), Via Bonomea 265, I--34136, Trieste, Italy}
\altaffiltext{2}{INAF-Osservatorio Astronomico di Padova, Vicolo dell'Osservatorio 5, I--35122, Padova, Italy}

\altaffiltext{$\star$}{Email: aatrani@gmail.com}



\begin{abstract}
Recent observations suggest ongoing planet formation in the innermost parsec of the Galactic center (GC). The super-massive black hole (SMBH) might strip planets or planetary embryos from their parent star, bringing them close enough to be tidally disrupted. Photoevaporation by the ultraviolet field of young stars, combined with ongoing tidal disruption, could enhance the near-infrared luminosity of such  starless planets, making their detection possible even with current facilities.  In this paper, we investigate the chance of planet tidal captures by means of high-accuracy N-body simulations exploiting Mikkola's algorithmic regularization. We consider both planets lying in the clockwise (CW) disk and planets initially bound to the S-stars. We show that tidally captured planets remain on orbits close to those of their parent star. Moreover, the semi-major axis of the planet orbit can be predicted by simple analytic assumptions in the case of prograde orbits. We find that starless planets that were initially bound to CW disk stars   have mild eccentricities and tend to remain in the CW disk. However, we speculate that angular momentum diffusion and scattering with other young stars in the CW disk might bring starless planets on low-angular momentum orbits. In contrast, planets initially bound to S-stars are captured by the SMBH on highly eccentric orbits, matching the orbital properties of the G1 and G2 clouds. Our predictions apply not only to planets but also to low-mass stars initially bound to the S-stars and tidally captured by the SMBH. 

\end{abstract}%

\bibliographystyle{apj}





\section{Introduction}
Several hundred young stars lie in the innermost parsec of our Galactic center (GC). The orbits of the so-called S-stars, $\sim 28$ young ($\approx{}20-100$ Myr) stars lying close ($< 0.04 \pc$) to the super-massive black hole (SMBH), provide the strongest constraints on the SMBH mass \citep{sch03,ghe03,gil09a}. The S-stars have been classified as B-type stars and have randomly oriented highly eccentric orbits. Hundreds of young ($\sim{}2-6$ Myr) stars  (mainly Wolf-Rayet and O-type stars, \citealt{paumard2006,lu2009,lu2013}) lie further out ($> 0.04 \pc$), $20\%$ of which form a nearly-Keplerian disk around the SMBH, named clockwise (CW) disk for its motion when projected on the plane of the sky \citep{bartko2009,yelda2014}. The formation mechanisms and dynamical evolution of such stars are still debated, since the tidal field of the SMBH is expected to disrupt molecular clouds in the innermost parsec  (e.g. \citealt{bonnell2008,hobbs2009,alig2011,alig2013,mapelli2008,mapelli2012,lucas2013,mapelli2016a,trani2016}, see \citealt{mapelli2016b} for a review).

Young stars in the local Universe are often surrounded by a protoplanetary disk (e.g. \citealt{williams2011} for a review). Thus, it is likely that protoplanetary disks exist even in the GC, despite the environment is quite hostile to star and planet formation. Indeed, recent radio continuum observations suggested the presence of photoevaporating protoplanetary disks in the innermost $\sim{}0.1$ pc \citep{yus15}. Whether planets can form in such protoplanetary disks is still highly uncertain.

\citet{map15} recently showed that starless planets are too faint to be observed in the GC with current facilities, even if they are photoevaporated by the intense ultraviolet (UV) emission of the young massive stars. However, if a  planet or protoplanetary embryo is undergoing tidal disruption by the SMBH field, the efficiency of photoevaporation can be enhanced by orders of magnitude: a Br-$\gamma$ luminosity of $\approx{}10^{31}$ erg s$^{-1}$ can be emitted in this case, observable with 10m-class telescopes \citep{map15}. Moreover, high-energy flares with a luminosity of $\leq{}2\times{}10^{41}$ erg s$^{-1}$ can be associated to tidal disruption events of planets by SMBHs \citep{zubovas2012}. The tidal disruption of smaller bodies, such as asteroids or planetesimals, is expected to be very frequent (although less energetic than that of planets), and has been invoked to explain the daily infrared flares of SgrA$^{\ast}$ \citep{cadez2008,kostic2009,zubovas2012,hamers2015}. 

Finally, a protoplanetary origin has been suggested even for the dusty object G2, which has been observed orbiting the SMBH on an highly eccentric orbit ($e \sim 0.98$) with extremely small pericenter ($a \sim 200 \AU$, \citealt{gil12,pfu15,wit14,gil13a,gil13b,eck13,phi13}). In fact, \citet{mur12} proposed that G2 is a low-mass star with a proto-planetary disk, while \citet{map15} suggested that the properties of G2 are consistent with a planetary embryo tidally captured by the SMBH.  The origin of G2 is still debated, and many other theories have been proposed to explain it: a gas cloud formed by colliding stellar winds \citep{bur12,sch12,gil13a,dec14,shc14} or tidally stripped material \citep{gui14}, the merger product of a binary \citep{pro15}, a low-mass star obscured by dust \citep{bal13,sco13,wit14}, a star disrupted by a stellar black hole \citep{mir12}, and a nova outburst \citep{mey12}.  Moreover, another similar object, named G1 \citep{ghe05,cle05}, has been suggested by \citet{pfu15} to share the same origin as G2. The orbit of G1 has lower eccentricity ($e \sim 0.93$) and smaller semi-major axis ($a \sim 2970 \AU$), but is very similar to that of G2.

In conclusion, whether planets and protoplanets exist in the GC is still an open question, and their detection with current facilities is challenging. Our aim is to study the dynamics of planets and protoplanets near the SMBH in the GC, in order to put constraints for future observations. In particular, we study the tidal capture of hypothetical planets and protoplanets  orbiting stars in the CW disk and in the S-star cluster. We simulate hierarchical three-body systems composed of a SMBH, a star, and a planet. In our three-body runs, the orbit of the star around the SMBH  is randomly sampled according to the properties of the CW disk. We also simulate the entire S-star cluster, adding a planet to each simulated S-star. In Section \ref{sec:methods} we describe the methodology we employed for our simulations; in Section \ref{sec:results} we present our results. In Section \ref{sec:discussion}, we discuss the implications of our work. Our conclusions are presented in Section \ref{sec:conclusions}.

\section{Methods}\label{sec:methods}
\subsection{Mikkola's Algorithmic Regularization code}
Modelling the evolution of planets close to the SMBH is challenging, because of the extreme mass ratios involved. Thus, our simulations are run by means 
of a fully regularized N-body code that implements the Mikkola's algorithmic regularization (MAR, \citealt{mikkola1999a,mikkola1999b}). This code is particularly suitable for studying the dynamical evolution of few-body systems in which strong gravitational encounters are very frequent and the mass ratio between the interacting objects is large. The MAR scheme removes the singularity of the two-body gravitational potential for $r\rightarrow{}0$, by means of a transformation of the time coordinate (see \citealt{mikkola1999a} for the details). 

Our implementation uses a leapfrog scheme in combination with the Bulirsh-Stoer extrapolation algorithm \citep{stoer1980} to increase the accuracy of the numerical results. 
The code integrates the equations of motion employing relative coordinates by means of the so called chain structure. This change of coordinates reduces round-off errors significantly \citep{aarseth2003}. At present, this code is a sub-module of the direct N-body code HiGPUs-R which
is still under development (Spera, in preparation; see \citealt{capuzzo2013}  for the current non-regularized version of HiGPUs). Still, it can be used as a stand-alone tool to study the dynamical evolution of few-body systems with very high precision.

Tidal dissipation is not taken into account in the current version of the code. In fact, we expect the effect of tidal dissipation to be negligible in our simulations, since the timescale of orbital decay is 
$\approx{}1\, \rm Gyr$, much longer than the length of our simulations ($10^3-10^4 \yr$).

\begin{figure*}[htbp]
	\begin{center}
		\includegraphics[width=0.497\linewidth]{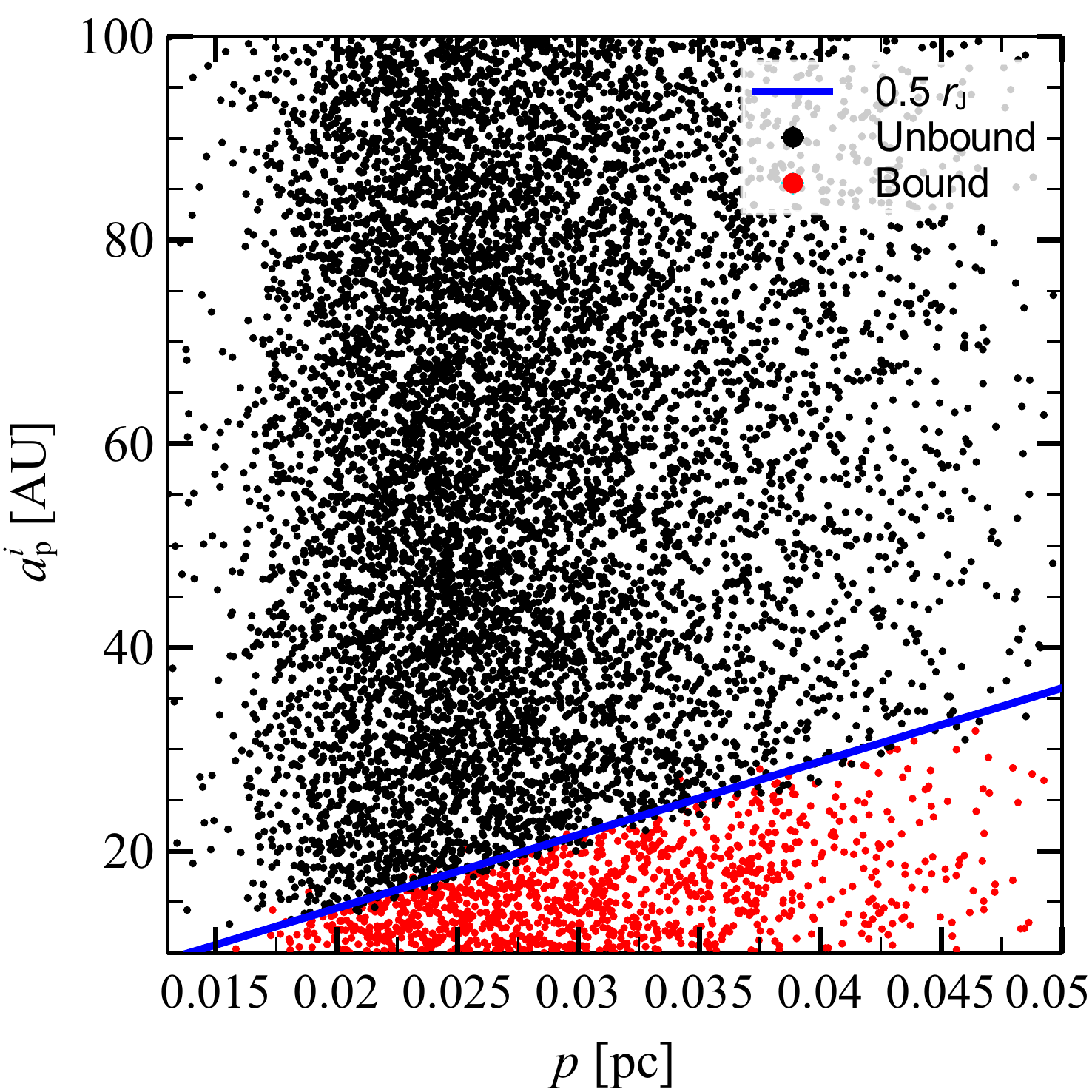}
		\includegraphics[width=0.497\linewidth]{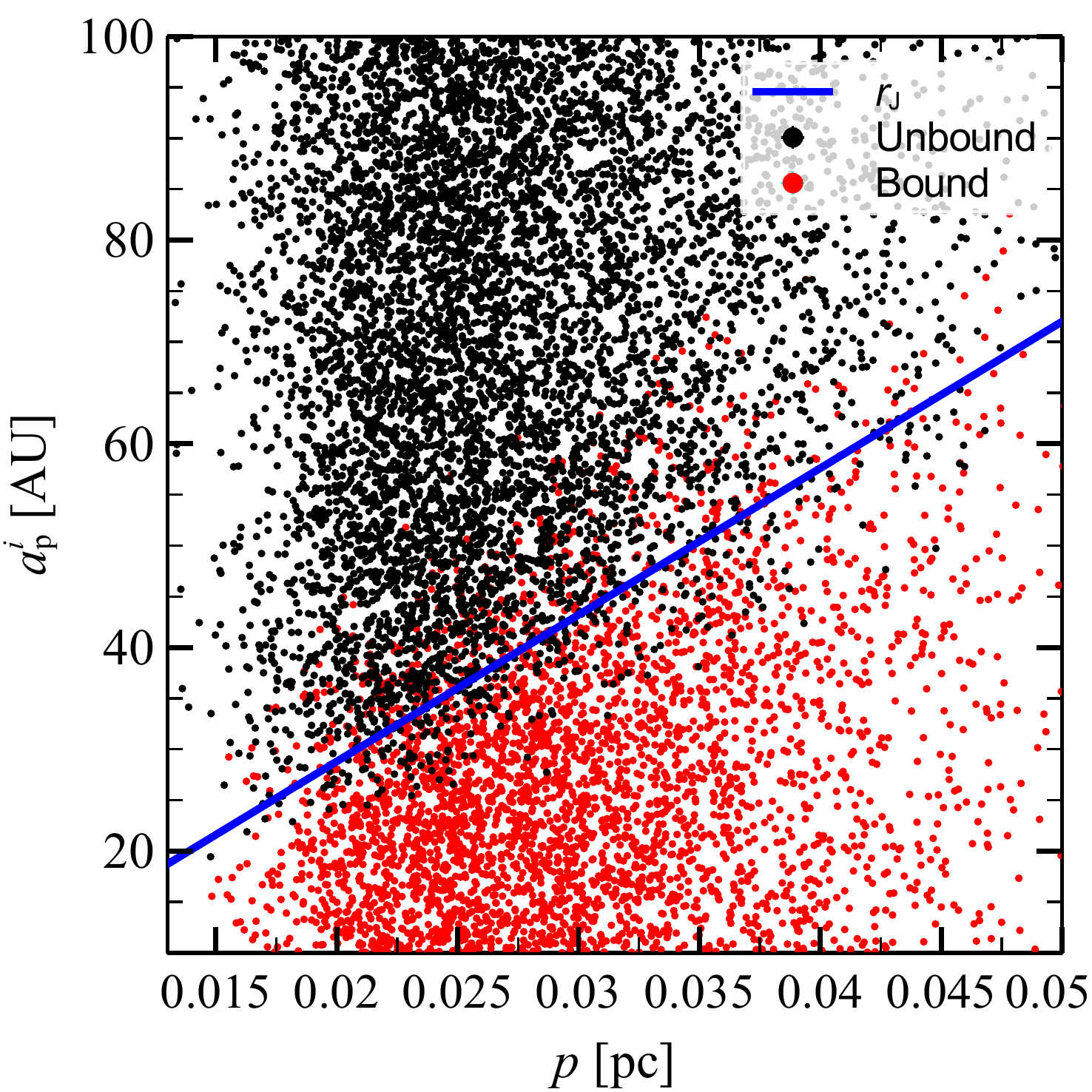}
		\caption{ \label{fig:unb} Initial semi-major axis of the planet versus pericenter distance of star orbit. Each dot represents a single realization of a three-body system of set~A (coplanar prograde, left-hand panel) and set~B (coplanar retrograde, right-hand panel). Red dots: realizations in which the planet remains bound to the star throughout the simulation. Black dots: realizations in which the planet becomes unbound with respect to the star. Blue solid line: Jacobi radius (Equation \ref{eq:rj}), multiplied by $0.5$ in the left-hand panel.%
		}
	\end{center}
\end{figure*}

\subsection{CW disk simulations}
Simulating the entire CW disk ($>1000$ stars) in the same run is prohibitive for MAR codes. Thus, we run simulations of a three-body hierarchical system composed of a SMBH, a star and a planet initially bound to the star.  We set the SMBH, star and planet masses to $4.31 \times 10^6 \msun$ \citep{gil09a}, $5 \msun$, and $10\mjup$, respectively, where $\mjup$ is the mass of Jupiter. 
The stellar orbit around the SMBH is modeled following the properties of the stars in the CW disk. The semi-major axis is drawn from a power-law distribution with index $\Gamma = 1.93$ \citep{do13}, in the range $0.03-0.06 \pc$, corresponding to the inner edge of the CW disk (planets orbiting CW stars on outer orbits are less likely affected by the SMBH tidal field). The star eccentricity is drawn from a Gaussian distribution centered at $0.3$ with $\sigma = 0.1$.

{ A planet will likely remain bound to the star if its distance from the star is less than Jacobi radius $r_{\rm J}$ of the star-planet system:
\begin{equation}\label{eq:rj}
r_{\rm J} = d \,\left(\frac{m}{3 M_{\rm SMBH}}\right)^{1/3},
\end{equation}
where $M_{\rm SMBH}$ is the SMBH mass, $m$ is the total mass of the star-planet system and $d$ is the distance between the star and the SMBH. 

With these initial conditions we expect the Jacobi radius to be in the range of  $20-90 \AU$ .
We assume that the planet orbit around the star is circular with radius in the uniform range $10-100 \AU$. Planets with a semi-major axis smaller than $10 \AU$ will unlikely be captured by the SMBH, while planets with semi-major axis larger than $100 \AU$ will be already unbound from the star.
We set the planet orbit eccentricity to zero in order to avoid the parameter space to explode. On the other hand, we expect that planets on eccentric orbits escape even faster.
 
We consider different inclinations with respect to the star orbit: coplanar prograde orbits ($i = 0^\circ$, set~A), coplanar retrograde orbits ($i = 180^\circ$, set~B), inclined prograde orbits (uniformly distributed over $270^\circ <i< 90^\circ$, set~C), and inclined retrograde orbits (uniformly distributed over $90^\circ <i< 270^\circ$, set~D). The mean anomalies of star and planet are uniformly distributed between 0 and 2$\pi$. We run $10^4$ realizations for each set and stop the simulations at $10^4 \yr$. Table \ref{tab:ic} shows a summary of the simulation sets presented in this paper. 
}
\begin{deluxetable}{clcc}
\tabletypesize{\scriptsize}
\tablecaption{Main properties of the simulations of planets in the CW disk.\label{tab:ic}}
\tablewidth{\linewidth}
\tablehead{
\colhead{Set} & \colhead{Planet orbit} & \colhead{$N$} & \colhead{$N_{\rm unb}$} 
}
\startdata
A & Coplanar, prograde & $10^4$ & $8903$ \\ 
B & Coplanar, retrograde & $10^4$ & $6488$ \\
C & Inclined, prograde & $10^4$ & $8817$ \\ 
D & Inclined, retrograde & $10^4$ & $7791$ 
\enddata

\tablecomments{\footnotesize 
Column~1: set name; column~2: planet orbit spin with respect to stellar orbit; column~3: number of realizations; column~4: number of realizations in which the planet becomes unbound with respect to the parent star.}
\end{deluxetable}

\subsection{S-star simulations}
Unlike the CW disk, the S-star cluster is sufficiently small to be simulated in the same run with the MAR algorithm. We run simulations of the 27 innermost S-stars for which the orbital elements are known, using as initial condition the orbital parameters reported by \citet{gil09a}. We assign to each star a planet of $10\mjup$ in circular orbit. { The planet semi-major axis ranges between $1$ and $20 \AU$, distributed in 20 equally spaced bins}. For each semi-major axis we run 1000 realizations randomizing the planet orbit orientation over the sphere, for a total of $20000$ realizations. We stop the simulations at $1000 \yr$.

\begin{figure*}[htbp]
	\begin{center}
		\includegraphics[width=0.497\linewidth]{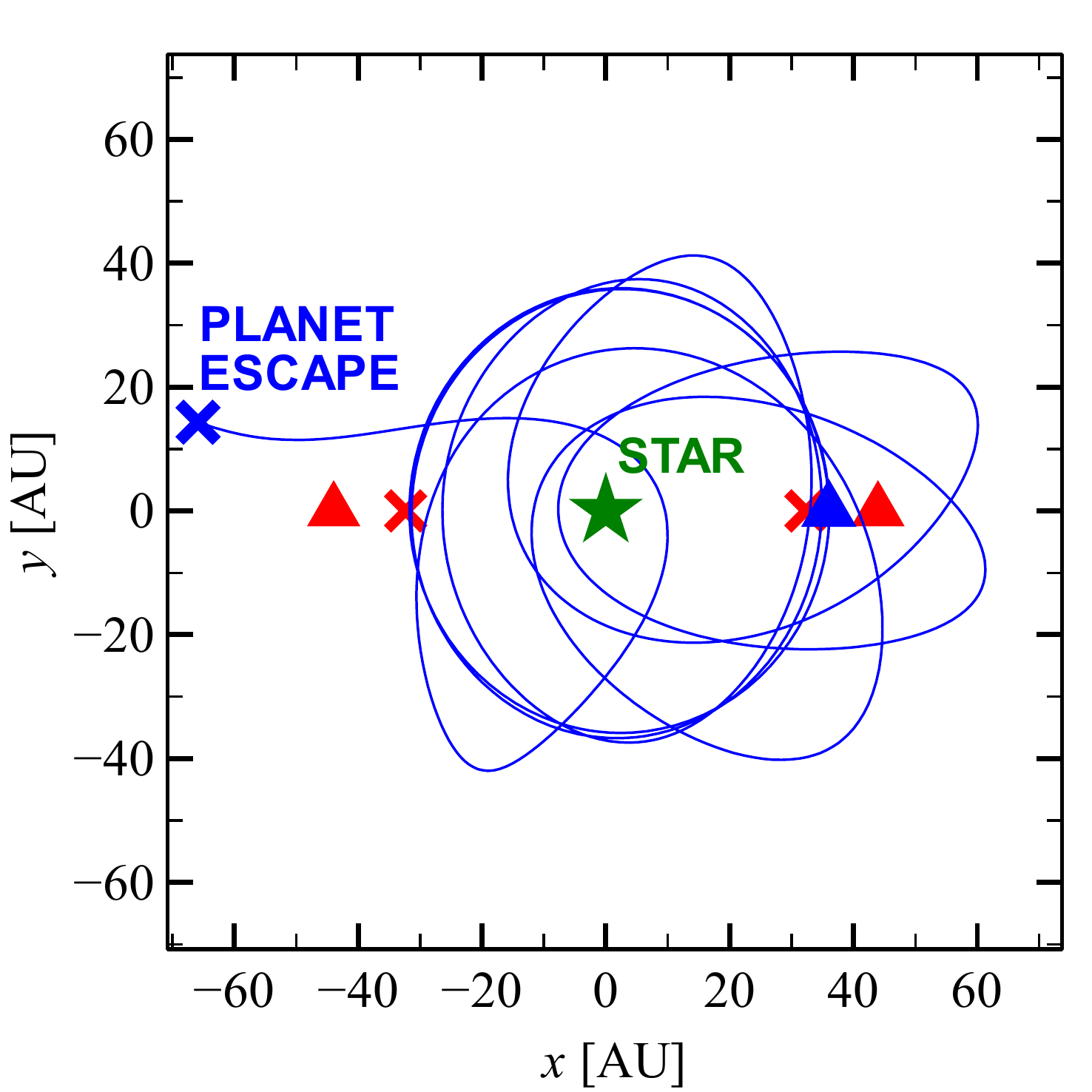}
		\includegraphics[width=0.497\linewidth]{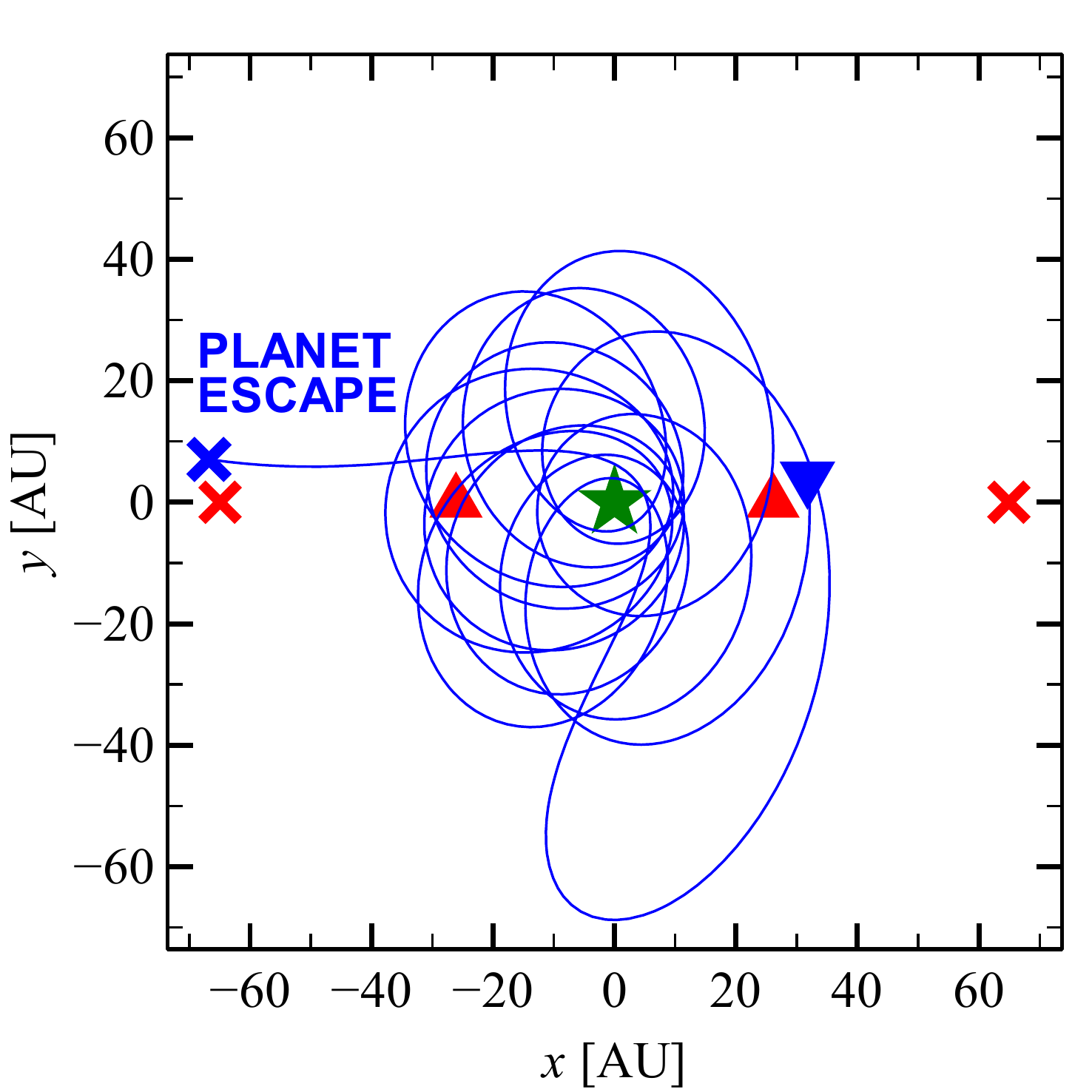}
		\caption{ \label{fig:orbital} Planet trajectory in the reference frame that corotates with the star for a single simulation of set~A (left-hand panel) and set~B (right-hand panel). The negative $x$-axis points always towards the SMBH, while the star tangential velocity lies in along the positive $y$-axis. Blue solid line: planet trajectory. Blue triangle: initial planet position. Blue cross: planet position at the time the planet becomes unbound with respect to the star. Green star: star position. Red triangle: initial Jacobi radius of the system (equation~\ref{eq:rj}), multiplied by $0.5$ in the left-hand panel. Red cross: same as red triangle, but at the time the planet becomes unbound with respect to the star.
		}
	\end{center}
\end{figure*}

\section{Results}\label{sec:results}
\subsection{Planets in the CW disk}
In $88-89 \%$ of the prograde runs (set~A~and~C) the planet escapes from the star and starts orbiting the SMBH. The escape fraction in retrograde runs (set~B~and~C) is lower: $65\%$ and $78\%$ of planets are tidally captured by the SMBH in set~B and set~D, respectively (see Table \ref{tab:ic}).
  
Figure~\ref{fig:unb} shows the initial semi-major axis of the planet $a^i_{\rm p}$ versus the pericenter distance $p$ of the stellar orbit for set~A (coplanar prograde, left-hand panel) and set~B (coplanar retrograde, right-hand panel). The colors indicate whether the planet remains bound to its parent star throughout the simulations.

The unbound and bound regions in the $a^i_{\rm p}-p$ plane are clearly distinct. The boundary between the two regions scales linearly with $p$, as expected from the Jacobi radius $r_{\rm J}$ linear dependence on the star-SMBH distance (equation~\ref{eq:rj}).

In the case of set~A (Figure \ref{fig:unb}, left-hand panel) the boundary is $0.5 \,{} r_{\rm J}$. In the case of set~B (Figure \ref{fig:unb}, left-hand panel) the boundary is $\sim{}1$ $r_{\rm J}$. The boundary is $\sim{}0.5$ $r_{\rm J}$ and $\sim{}0.9$ $r_{\rm J}$ for set~C~and~D, respectively.
{
Thus, the boundary radius  is smaller for prograde orbits than for retrograde orbits. This difference is connected with the direction of the Coriolis force. Moreover, the boundary is less sharp in the case of retrograde orbits. This likely occurs because retrograde planets spend several periods at radius $\sim r_{\rm J}$ without escaping, thanks to the stabilizing effect of the Coriolis force. In contrast, prograde planets escape immediately outside $0.5 r_{\rm J}$. As a consequence, planets in retrograde orbits are more affected by perturbations from the tidal field, which is stronger at larger distances from the star \citep{ham91,ham92}.
}

\begin{figure*}[htbp]
	\begin{center}
		\includegraphics[width=0.497\linewidth]{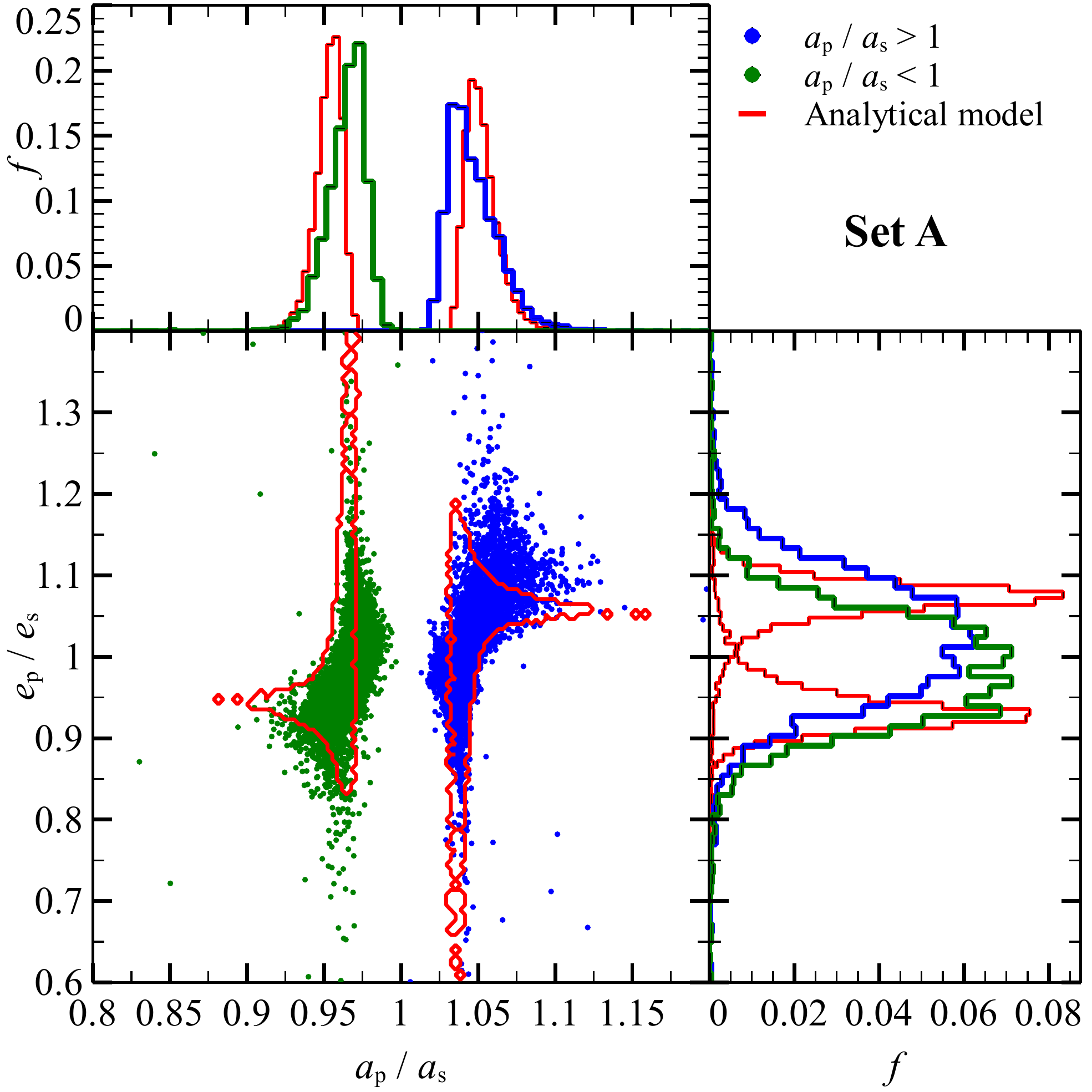}
		\includegraphics[width=0.497\linewidth]{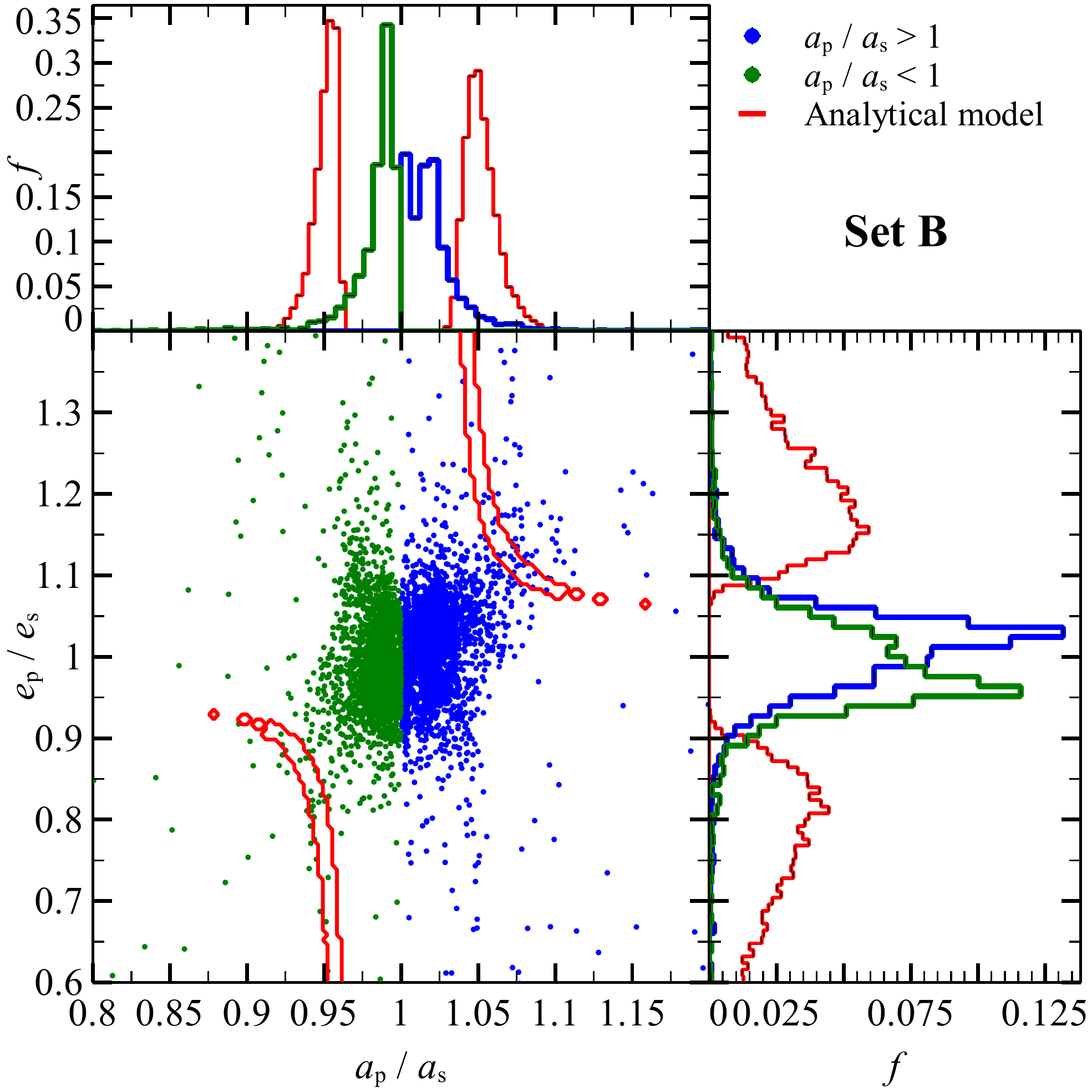}
        \includegraphics[width=0.497\linewidth]{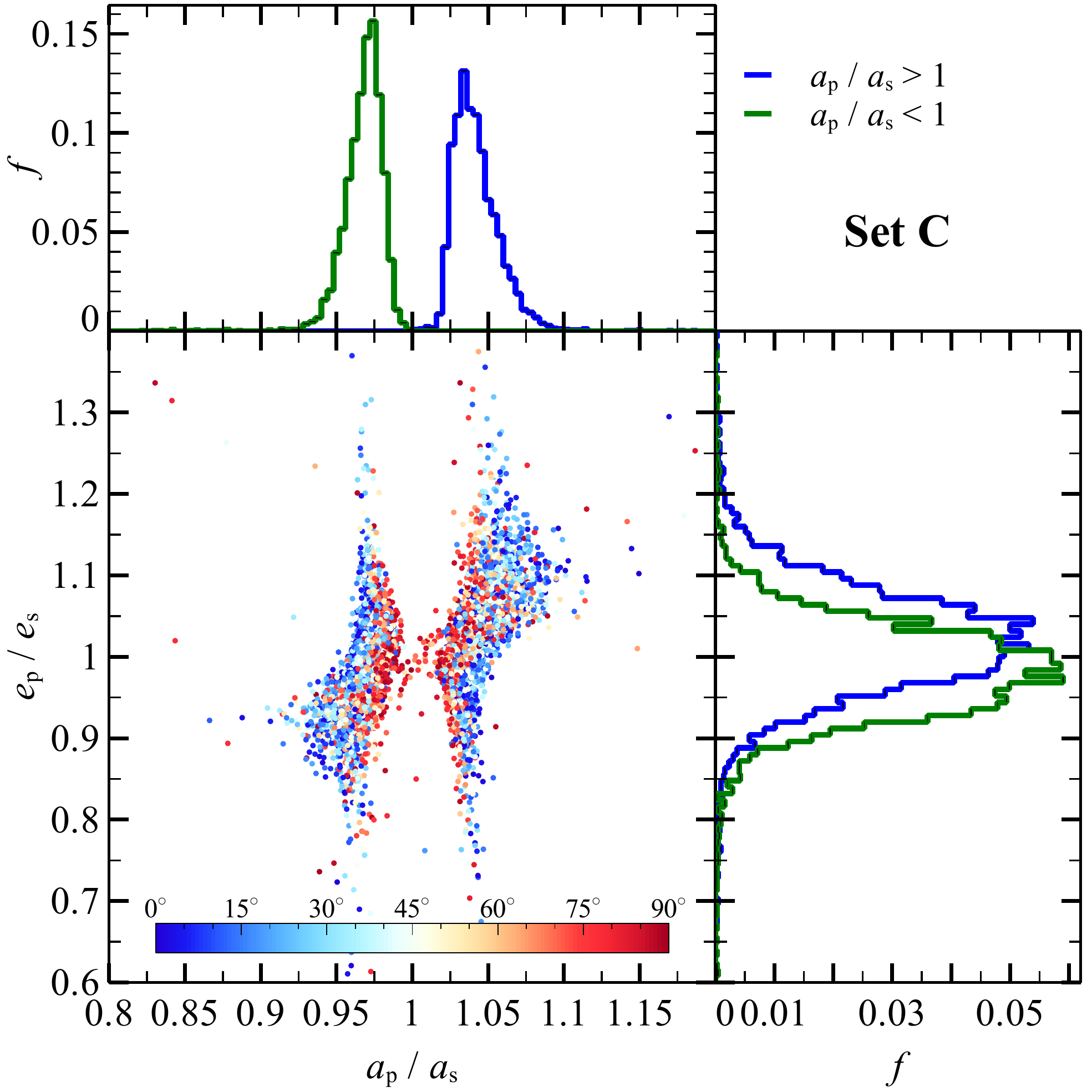}
		\includegraphics[width=0.497\linewidth]{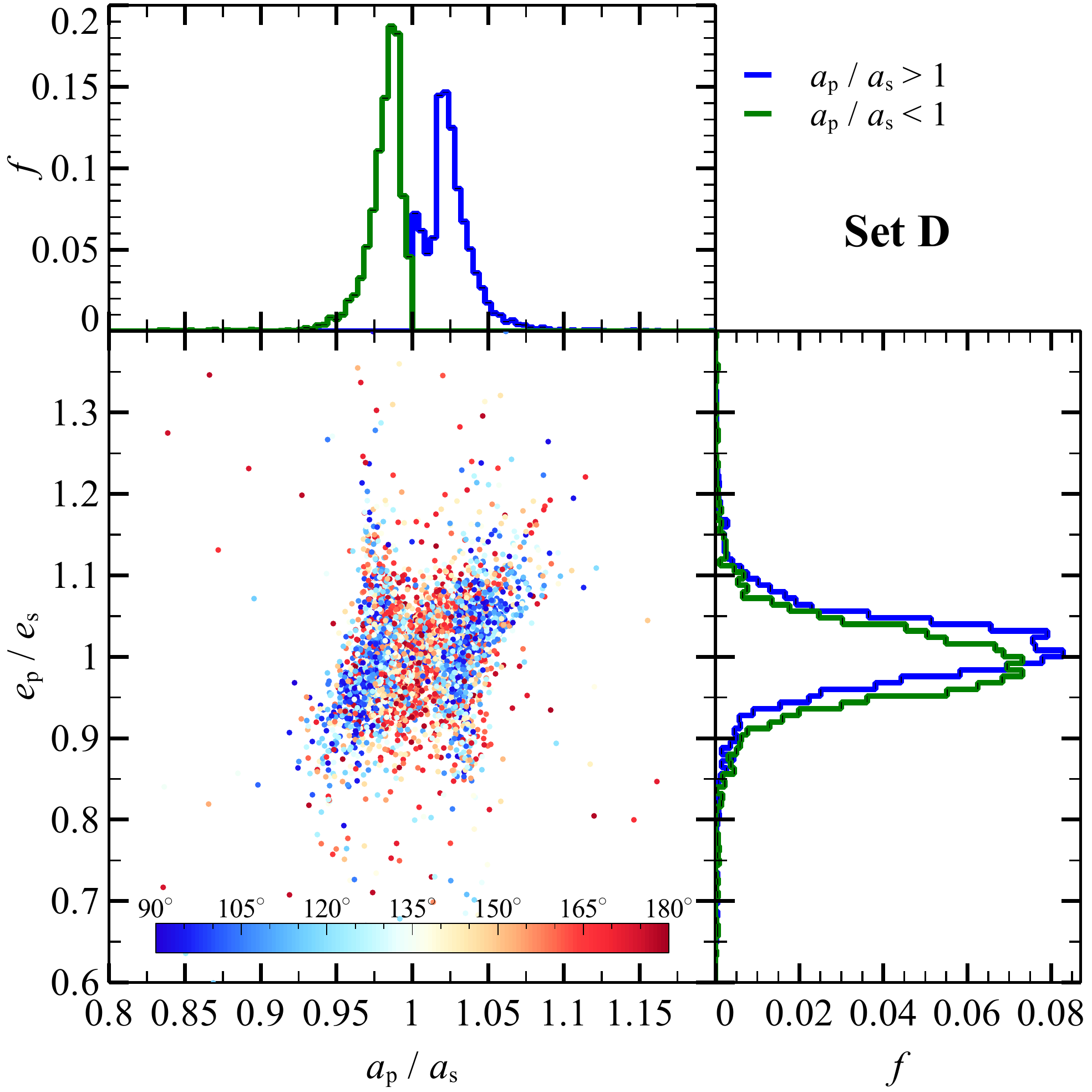}
		\caption{ \label{fig:ratios} Semi-major axis of the planet orbit (around the SMBH) normalized to the semi-major axis of the star orbit $a_{\rm p}/a_{\rm s}$ versus eccentricity of the planet orbit (around the SMBH) normalized to the eccentricity of the parent star orbit $e_{\rm p}/e_{\rm s}$. In the top panels: blue dots indicate realizations in which the planet semi-major axis is larger than its parent star semi-major axis ($a_{\rm p}/a_{\rm s} > 1$); green dots indicate realizations in which the planet semi-major axis is smaller than its parent star semi-major axis ($a_{\rm p}/a_{\rm s} < 1$); red contours indicate predictions of the analytic model (equations \ref{eq:deltas}). In the bottom panels: the color map indicates the inclination of the planet orbit with respect to the star orbit. $i = 0^\circ, 90^\circ$ and $180^\circ$ correspond to prograde coplanar orbits, normal orbits and retrograde coplanar orbits, respectively.
                  In all panels: green histograms indicate the distributions of planets with $a_{\rm p}/a_{\rm s} < 1$, while blue histograms indicate the distributions of planets with $a_{\rm p}/a_{\rm s} > 1$.
			Top-left panel: set~A (coplanar prograde runs). Top-right panel: set~B (coplanar retrograde runs). Bottom-left panel: set~C (inclined prograde runs). Top-right panel: set~D (inclined retrograde runs).%
		}
	\end{center}
\end{figure*}

Figure~\ref{fig:orbital} shows the trajectory of a planet in a single simulation of set~A (coplanar and prograde, left-hand panel) and set~B (coplanar and retrograde, right-hand panel). The reference frame corotates with the star in its motion around the SMBH, so that the SMBH is always directed towards the negative $x-$axis. 
In the left-hand panel, the planet orbit is initially within half of the Jacobi radius and the planet completes an orbit around the star before being captured by the SMBH. However, as the star moves towards its pericenter, the Jacobi radius of the system shrinks and the planet is captured by the tidal forces of the SMBH.

In the case of retrograde orbits (Figure \ref{fig:orbital}, right-hand panel), the planet trajectory can be much more convoluted. In this case, the planet orbit becomes unstable after the third pericenter passage of the parent star around the SMBH; the orbit of the planet becomes prograde before escaping from the Hill sphere of the star. Moreover, the Hill sphere at the initial time is smaller than that at the moment of planet escape, indicating that planet escape does not occur at pericenter passage.
{
For more details about the temporary orbit of simulated planets see Appendix~\ref{sec:app}.
}

Figure \ref{fig:ratios} shows the orbital properties of the planets after they are captured by the SMBH. In $95\%$ of the runs of set~A (Figure \ref{fig:ratios}, top-left panel) the semi-major axis of the planet $a_{\rm p}$ (with respect to the SMBH) differs less than $7\%$ from the semi-major axis of its parent star $a_{\rm s}$.

The small difference between $a_{\rm p}$ and  $a_{\rm s}$ is motivated by the change of the orbital energy of the planet being of the same order of magnitude as the  binding energy of the star-planet system ($E_{\rm sp}$). For our assumptions, $E_{\rm sp}\approx{}10^{43}$ erg s$^{-1}$. This is much smaller than the binding energy between the star and the SMBH ($\approx{}10^{49}$ erg s$^{-1}$), indicating that the recoil velocity acquired by the planet during the capture event is much smaller than its initial velocity with respect to the SMBH.

The gap in the semi-major axis distribution in the top-left panel of Figure \ref{fig:ratios} indicates that the semi-major axis of the escaped planet is never equal to the semi-major axis of the parent star. The gap becomes wider as the planet orbit eccentricity deviates from that of its parent star.

The eccentricity distribution depends on whether the planet has a smaller or larger semi-major axis than its parent star. In the case of smaller semi-major axis, the eccentricity distribution is centered at lower eccentricity relative to the parent star, while in the case of larger semi-major axis the eccentricity distribution is centered at higher eccentricity relative to the parent star. 
In $95\%$ of the runs of set~A the eccentricity of the planet orbit $e_{\rm p}$ differs less than $15\%$ from the eccentricity of its parent star $e_{\rm s}$.

In runs of set~B (retrograde coplanar runs, see Figure \ref{fig:ratios}, top-right panel), the distribution of semi-major axis of planets normalized to that of stars with respect to the SMBH ($a_{\rm p}/a_{\rm s}$) has no gaps. The spread in semi-major axis is lower than in set~A, while the spread of eccentricities is higher. As in set~A, tighter planet orbits tend to have higher eccentricity and {\it vice versa.}

The bottom panels of Figure \ref{fig:ratios} show the orbital properties of the planets for the runs of set~C (inclined and prograde, bottom-left panel) and set~D (inclined and retrograde, bottom-right panel). Inclined orbits follow the same trend as coplanar ones: runs of set~C exhibit a gap in the $a_{\rm p}/a_{\rm s}$ distribution, while runs of set~D show no gap.

{ About 51\% runs of set~A~and~C (prograde runs) have $a_{\rm p}<a_{\rm s}$. In contrast, just $45\%$ and $43\%$ runs of set~B~and~D (retrograde runs) have $a_{\rm p}<a_{\rm s}$, respectively. In the retrograde runs, the planets tend to end on orbits less bound than those of their parent star. }

\begin{figure*}[htbp]
	\begin{center}
		\includegraphics[width=0.497\linewidth]{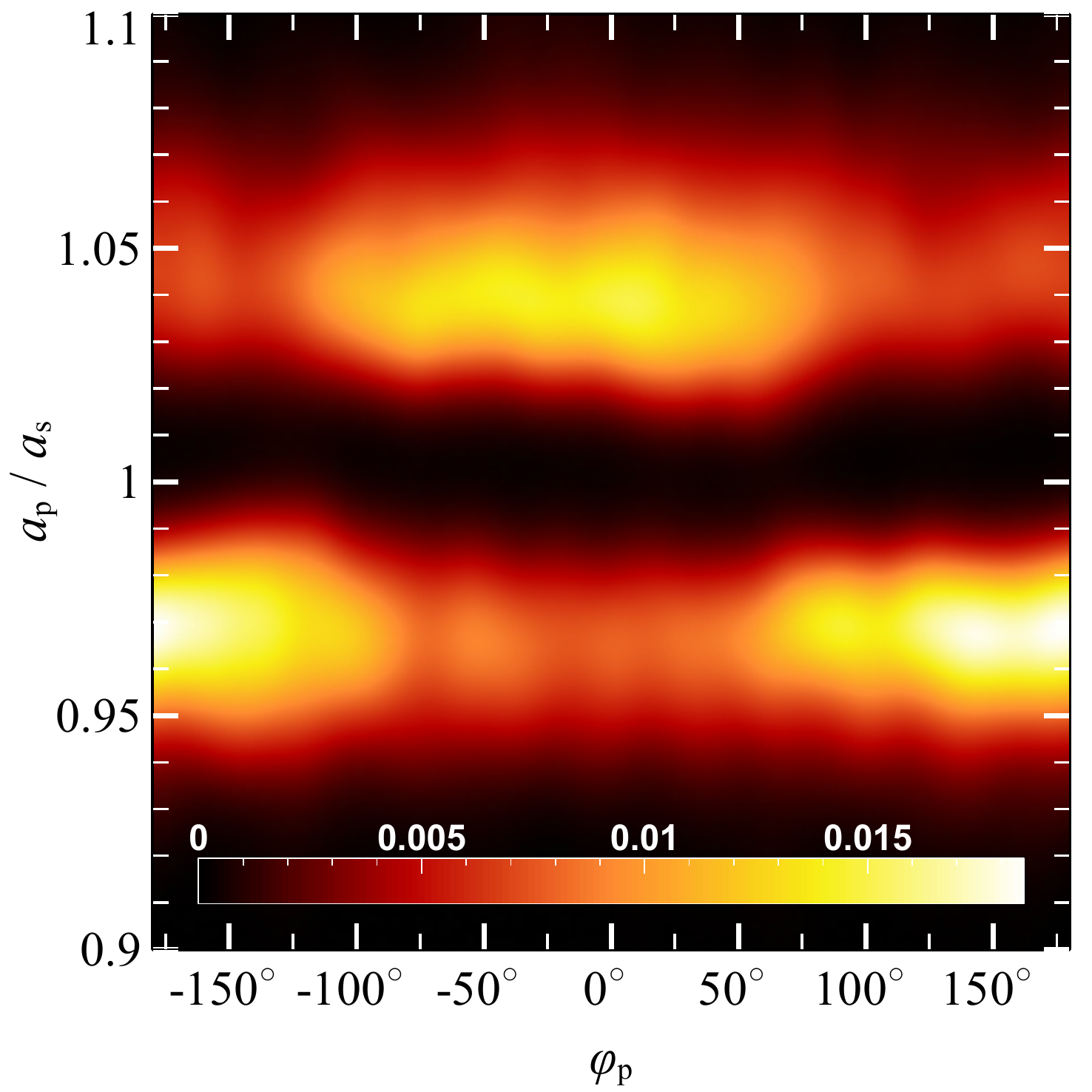}
		\includegraphics[width=0.497\linewidth]{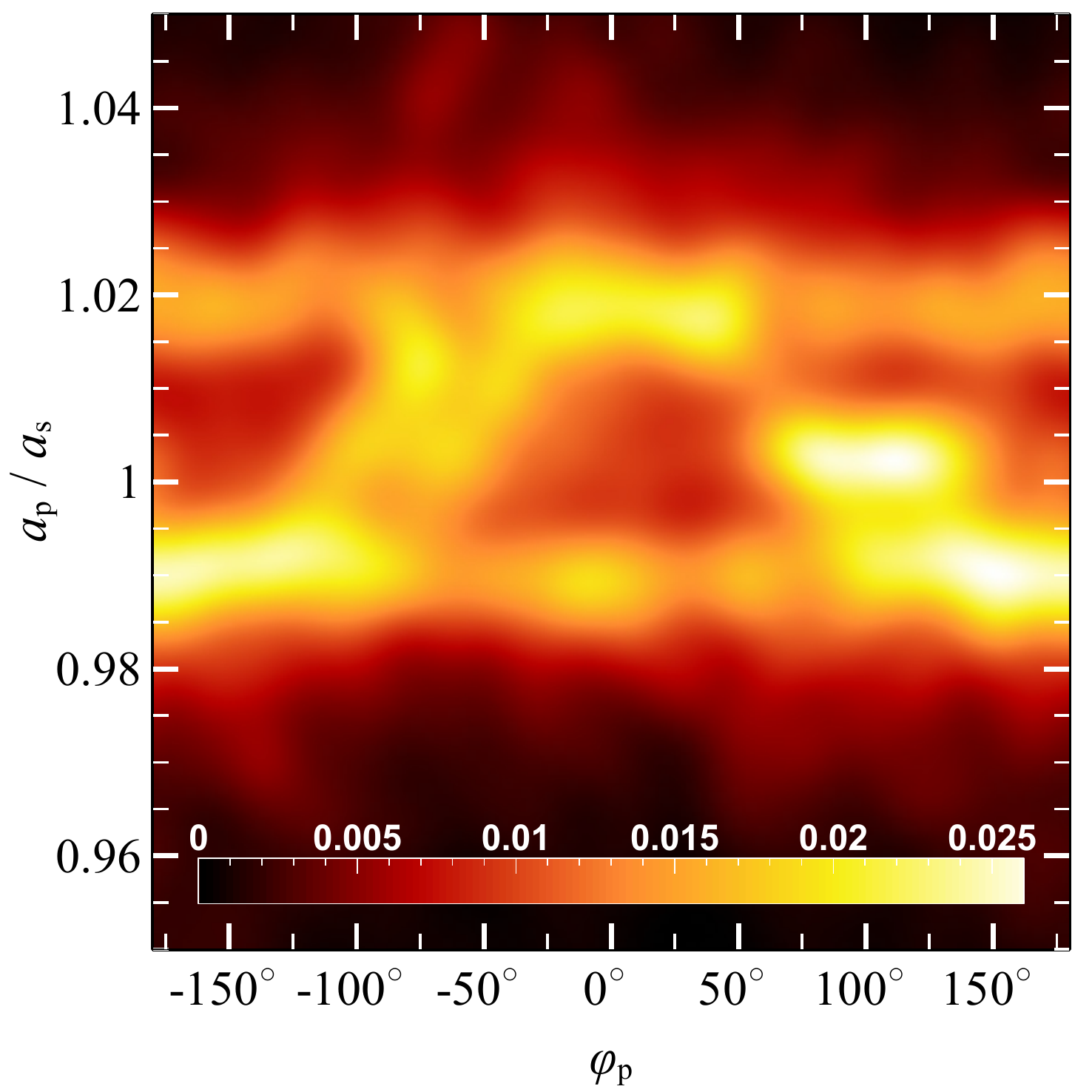}
		\caption{ \label{fig:aph} Probability density map of the ratio of the planet semi-major axis $a_{\rm p}$  to that of the star $a_{\rm s}$ versus the planet orbital phase around the star at the star first pericenter passage. $\varphi_{\rm p} = 180^\circ$ indicates that the planet is in between the SMBH and the star, while $\varphi_{\rm p} = 0^\circ$ indicates that the planet is in opposition with respect to the SMBH. Left-hand panel: set~A (coplanar prograde runs). Right-hand panel: set~B (coplanar retrograde runs).%
		}
	\end{center}
\end{figure*}

\begin{figure}[htbp]
	\begin{center}
		\includegraphics[width=1\columnwidth]{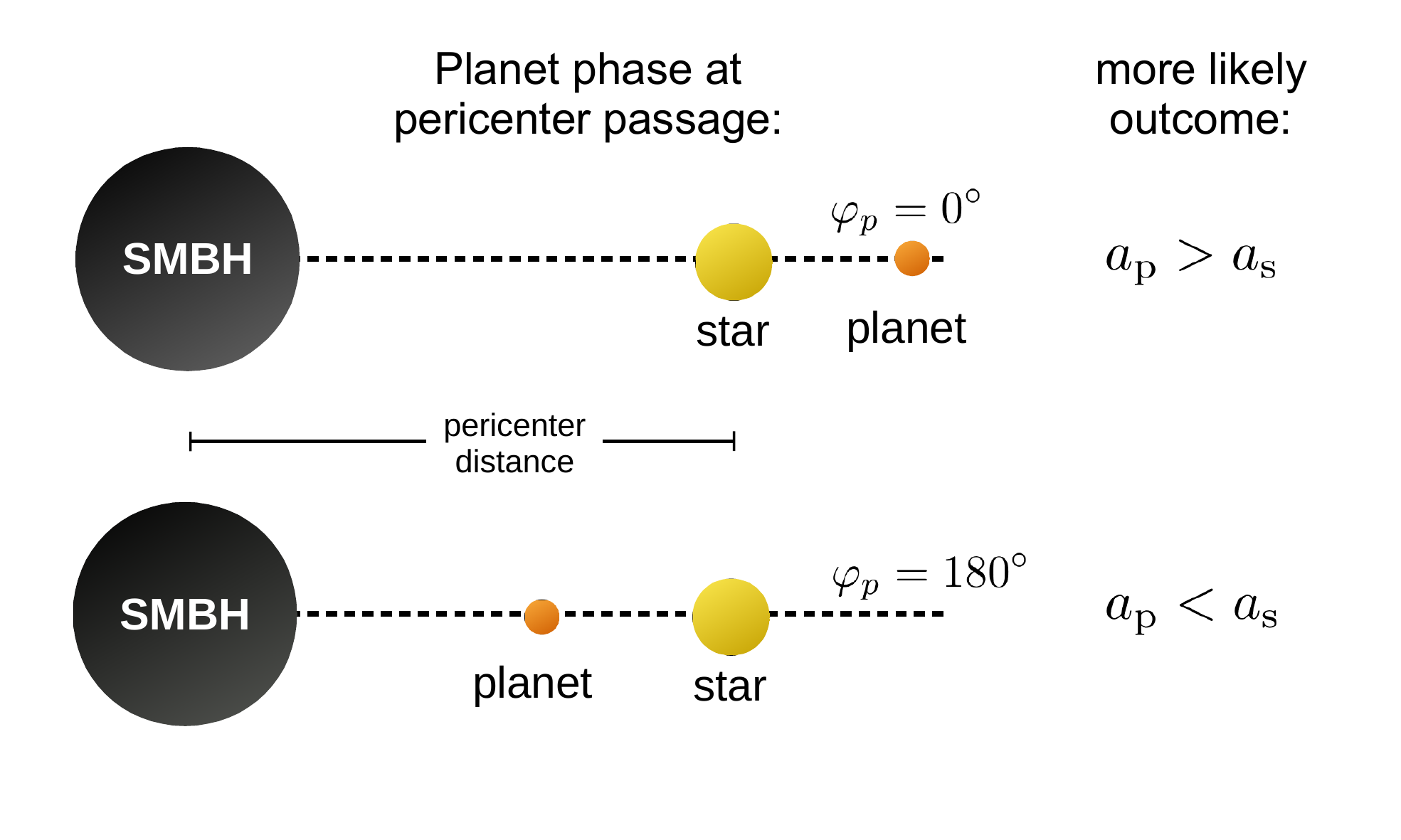}
		\caption{\label{fig:cartoon}  Schematic representation of two extreme orbital phases of the planet at the star pericenter passage, along with the more likely outcomes if the planet gets stripped from its parent star. $a_{\rm p}$: planet semi-major axis with respect to the SMBH after it becomes unbound, $a_{\rm s}$: parent star semi-major axis, $\varphi$: planet orbital phase at the star pericenter passage.%
		}
	\end{center}
\end{figure}

Figure \ref{fig:aph} shows the ratio $a_{\rm p}/a_{\rm s}$ between the planet semi-major axis and that of its parent star versus the orbital phase of the planet at the first pericenter passage of the star. We predict the orbital phase analytically using the initial conditions of each realization. From the left-hand panel of Figure \ref{fig:aph} it is apparent that the planet will likely have a semi-major axis smaller than that of its parent star ($a_{\rm p} / a_{\rm s} < 1$) in runs of set~A  if it is in between the SMBH and the star during the stellar pericenter passage ($\varphi_{\rm p} \simeq 180^\circ$). In contrast, the planet will likely have a semi-major axis larger than that of its parent star ($a_{\rm p} / a_{\rm s} > 1$) if the planet is on the opposite side of the orbit with respect to the SMBH ($\varphi_{\rm p} \simeq 0^\circ$). Figure \ref{fig:cartoon} is a schematic representation of this result.  The same trend is still present (but much  less evident) in runs of set~B (Figure \ref{fig:aph}, right-hand panel). 

We find that the planet may undergo a close encounter with the star during its orbit around the SMBH. This occurs because the planet remains on a orbit similar to that of its parent star, so that it may encounter again the star after one synodic period. However, since the difference between the orbital periods of the star and the planet is negligible, the synodic period is $\gtrsim 5000 \yr$. On this timescale, perturbations from nearby stars might become non-negligible before the planet undergo the encounter with its parent star. 

\subsection{Planets in S-stars cluster}
\begin{figure}[htbp]
	\begin{center}
		\includegraphics[width=1\columnwidth]{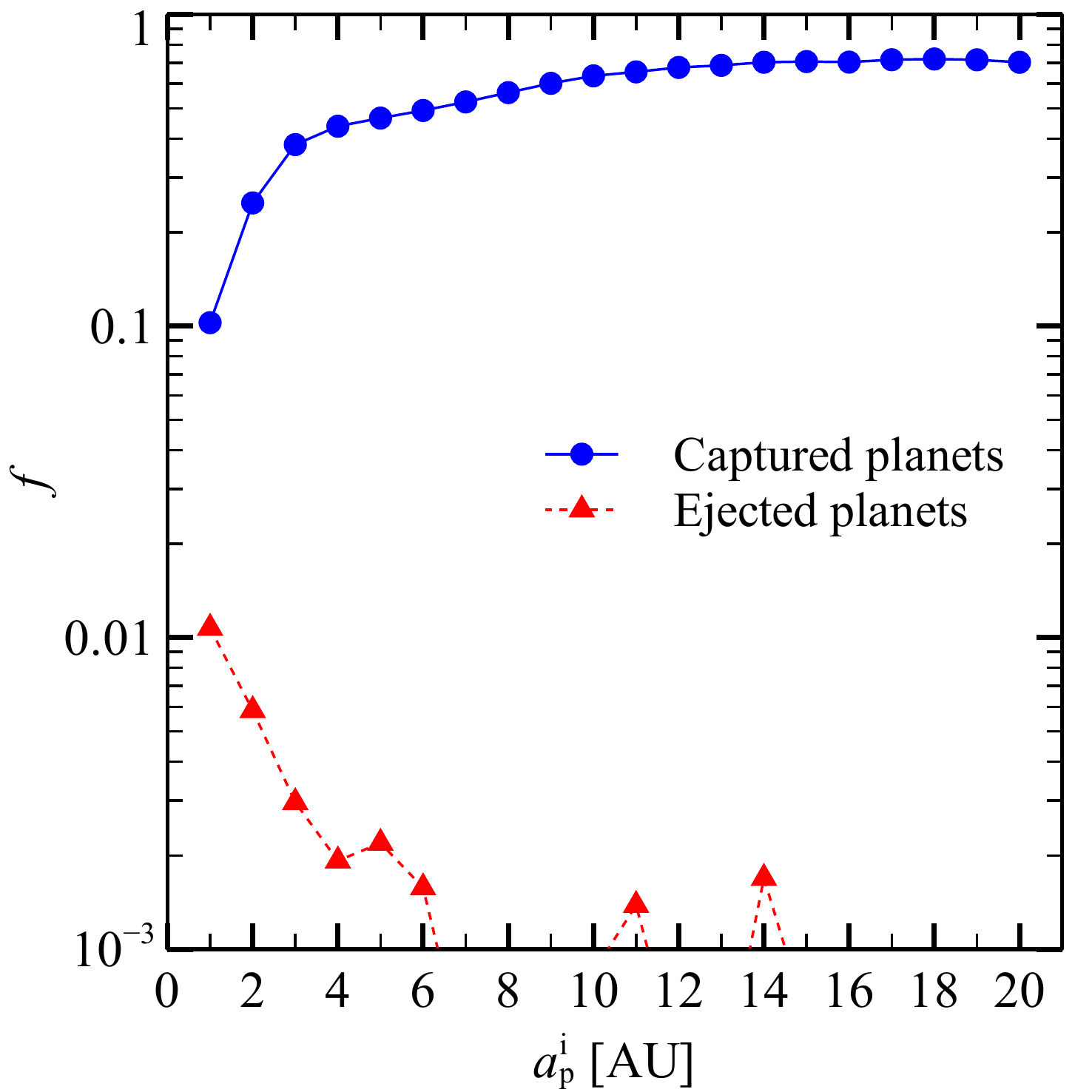}
		\caption{ \label{fig:ssfrac} Fraction of captured and ejected planets as function of initial semi-major axis of planet orbit for all S-stars realizations. Blue solid line: fraction of unbound planets. Red dashed line: fraction of planets ejected from the system. Planets whose initial semi-major axis is larger than the Jacobi radius of the star at the initial conditions are not included in this Figure.
		}
	\end{center}
\end{figure}

Figure \ref{fig:ssfrac} shows the fraction of captured and ejected planets versus the initial semi-major axis of planet orbits for all S-stars realizations. As expected, the fraction of unbound planets increases with the initial semi-major axis. $57 \%$ of the planets in our simulations gets captured by the SMBH. 
In total $0.18 \%$ of the planets get ejected from the system. The fraction of ejected planets decreases for larger initial semi-major axis. This is expected: the larger the semi-major axis, the smaller the binding energy of the planet-star system that can be released as recoil velocity during the encounter with the SMBH.

Figure \ref{fig:3dplanet} shows the trajectory of a planet around S19 star, in the rotating reference frame that corotates with the star. The star orbit lies in the $x$-$y$ plane, and the negative $x$-axis is always directed towards the SMBH. The planet orbit has an initial radius of $10 \AU$ and it is inclined by $20^\circ$ with respect to the star orbit. The planet orbit becomes immediately eccentric ($e \simeq 0.8$) due to the strong tidal forces and gets an inclination by $45^\circ$ and a semi-major axis of $8 \AU$. The orbit remains stable around the star for several periods, until the planet is kicked into a looser orbit with $i \simeq 100^\circ$, $a \simeq 20 \AU$ and $e \simeq 0.3$. After $260 \yr$, the planet escapes along the negative $x$-axis and gets captured by the SMBH.

The morphology of planet orbits varies greatly from simulation to simulation. Flips of planet orbit may occur, with the planet spending time on several temporarily-stable orbits around the star before escaping. In Section~\ref{sec:clouds} we compare the orbital parameters of the captured planets with the ones of the G1 and G2 cloud.

\section{Discussion}\label{sec:discussion}

\subsection{Orbital properties of unbound planets}


{ As shown in Figure~\ref{fig:ratios}, planets remain on orbits similar to those of their parent star after being captured by the SMBH. This implies that the velocity kick induced by the SMBH is at least one order of magnitude less than the star orbital velocity.
Furthermore, there is a gap in the distribution of the semi-major axes of captured planets in the prograde case.

Figure~\ref{fig:aph} (showing the semi-axis ratio $a_{\rm p}/a_{\rm s}$ versus the orbital phase of the planet) suggests that planets escaping from L1 (inner Lagrangian point) end on tighter orbits, while planets escaping from L2 (outer Lagrangian point) end on looser orbits.


\begin{figure}[htbp]
	\begin{center}
		\includegraphics[width=1\columnwidth]{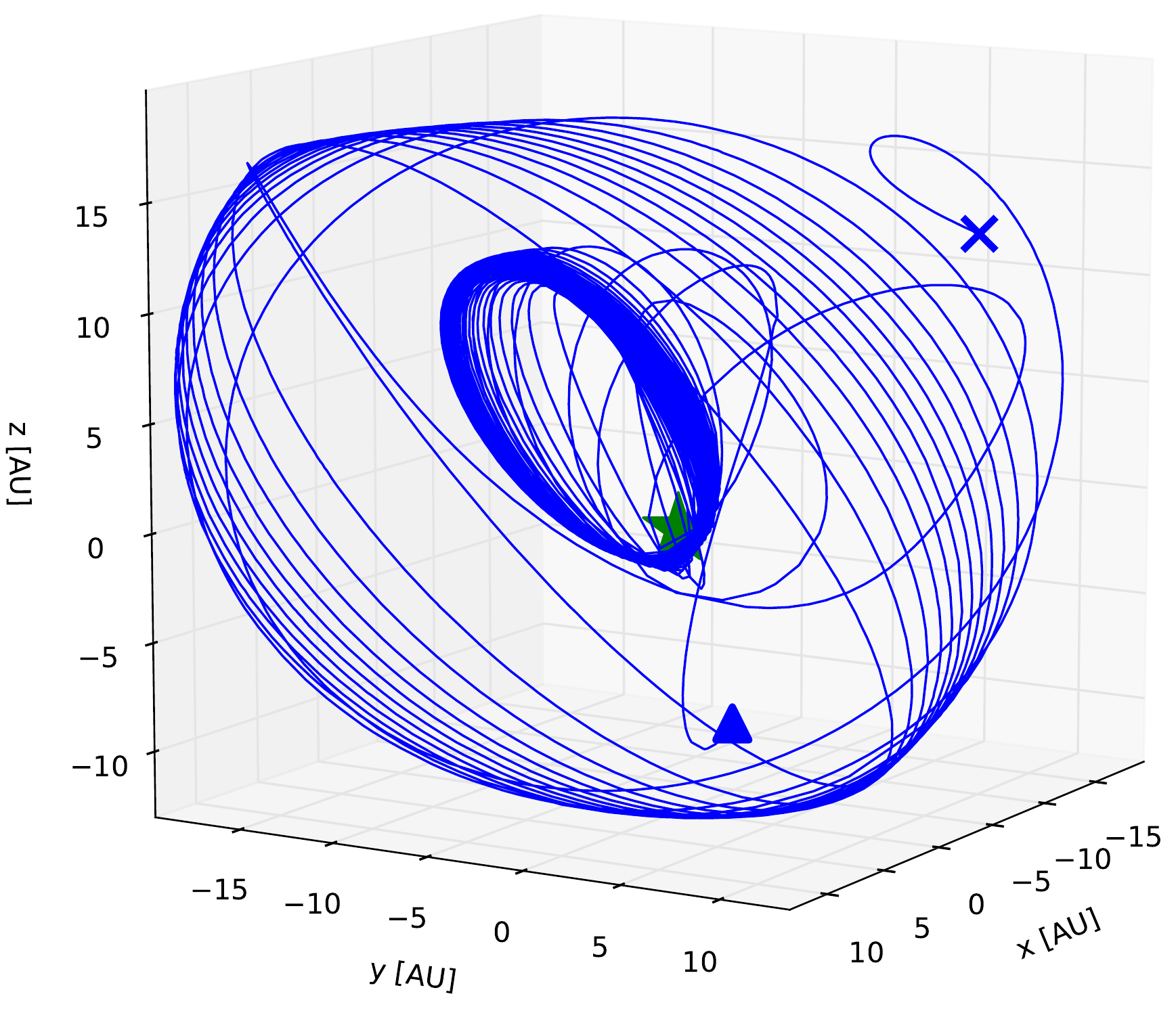}
		\caption{\label{fig:3dplanet} Planet trajectory around the star S19 of the S-star cluster, in the  reference frame that corotates with the star. The initial semi-major axis of the planet is $10 \AU$. Blue triangle: initial planet position. Blue cross: planet position at the time the planet becomes unbound with respect to the star ($540 \yr$). Green star: star position. The SMBH is located along the negative $x$-axis, while the star tangential velocity is directed along the positive $y$-axis.
		}
	\end{center}
\end{figure}

\begin{figure*}[htbp]
	\begin{center}
		\includegraphics[width=0.497\linewidth]{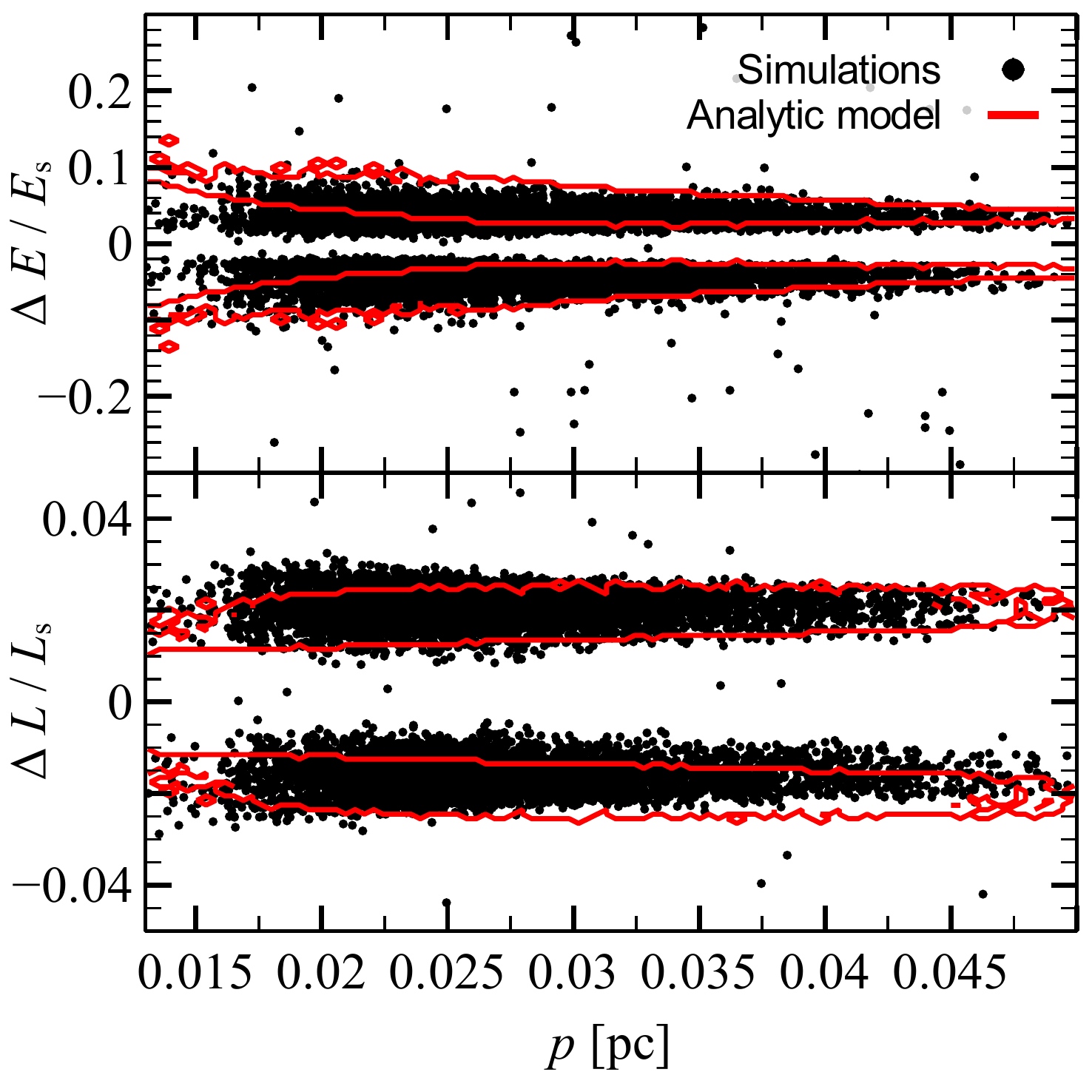}
		\includegraphics[width=0.497\linewidth]{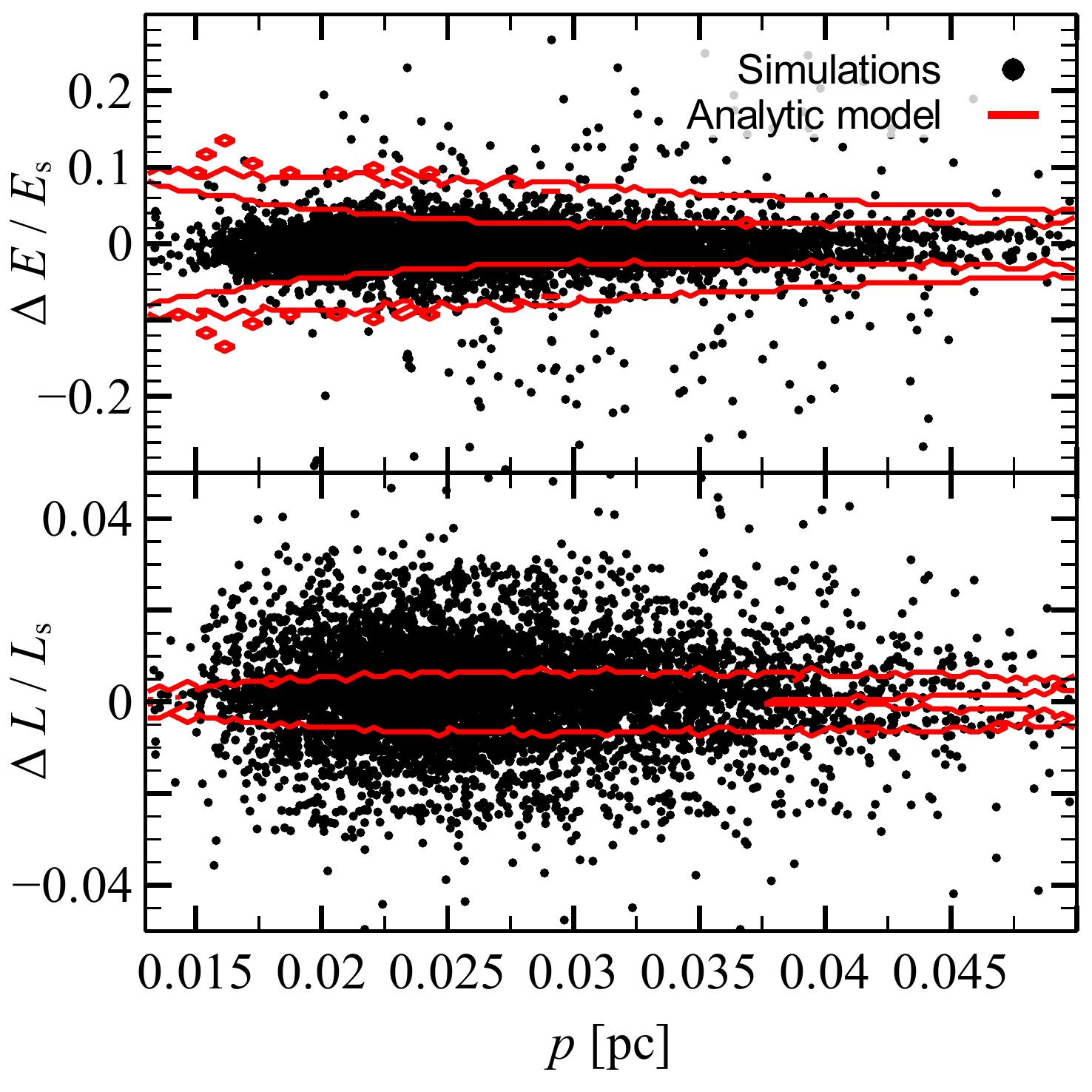}
		\caption{ \label{fig:deltas} Top (bottom) panels: energy (angular momentum) difference between planet and star orbits around the SMBH as a function of the pericenter distance of the stellar orbit, normalized to the star energy (angular momentum). Black dots: { results of} the simulations. Red contours: predictions of the analytic model (equations \ref{eq:deltas}). Left-hand panel: set~A (coplanar prograde runs). Right-hand panel: set~B (coplanar retrograde runs).%
		}
	\end{center}
\end{figure*}

 Based on these considerations}, we can estimate the change in specific angular momentum and energy of the planet in the framework of the restricted three-body problem.
We develop a simple analytic model based on three assumptions: (i) the planet becomes unbound during the star pericenter passage, (ii) the planet escapes the Hill sphere of the star from either the outer or the inner Lagrangian point, (iii) the planet velocity with respect to the rotating frame of reference at the moment of escape equals its orbital velocity $v_{\rm p}$.
With these assumptions we can compute the difference of the specific energy and angular momentum between the planet and the star orbit, $\Delta E$ and $\Delta L$, respectively:
\begin{equation}\label{eq:deltas}
\begin{aligned}
& \Delta E = - \frac{G\,{}M_{\rm SMBH}}{p} \frac{r_{\rm J}}{p - r_{\rm J}} - v_{\rm s}^2 \frac{r_{\rm J}}{p} \left(1 - \frac{1}{2} \frac{r_{\rm J}}{p}\right) \\ 
& \Delta L = - r_{\rm J}\,{} v_{\rm s} - p\,{} v_{\rm p} + r_{\rm J}\,{} v_{\rm p},  \\
\end{aligned}
\end{equation}
where $G$ is the gravitational constant, $M_{\rm SMBH}$ is the SMBH mass, $p$ is the pericenter distance of the star orbit, $r_{\rm J}$ is the Jacobi radius at pericenter (equation~\ref{eq:rj}), $v_{\rm s}$ is the star velocity at pericenter, and $v_{\rm p}$ is the orbital velocity of the planet. The sign of $r_{\rm J}$ and $v_{\rm p}$ is positive if the planet escapes from the inner Lagrangian point, negative if the planet escapes from the outer Lagrangian point, and $v_{\rm p}$ changes sign for retrograde orbits.

{
Figure~\ref{fig:deltas} shows the variation of energy ($\Delta E$) and angular momentum ($\Delta L$) predicted from the analytic model compared to the simulations. In the case of prograde orbits (left-hand panel, set~A), the simple analytic model reproduces very well the bimodal energy distribution.
The analytic model overestimates $\Delta E$ with decreasing pericenter distance, because the planet may escape before reaching the pericenter, if the pericenter is very small. In contrast, the analytic model does not match  the variation of energy and angular momentum in the simulations with retrograde orbits (right-hand panel of Figure~\ref{fig:deltas}, set~B).

Inserting the values drawn from the initial conditions of our simulations into equation~\ref{eq:rj} and \ref{eq:deltas} we can evaluate the parameters of the planet new orbit around the SMBH. In Figure \ref{fig:ratios} we plot the predicted  $a_{\rm p}/a_{\rm s}$ and $e_{\rm p}/e_{\rm s}$ along with the results of the simulations.

The predicted semi-major axis distribution matches the simulations in the case of prograde orbits (set~A, Figure \ref{fig:ratios}, left-hand panel), reproducing  the gap in the semi-major axis distribution. 

However, the analytic model also predicts a bimodality in the eccentricity distribution, which is not present in the simulations. In particular, the analytic model predicts that tighter orbits have mostly lower eccentricity and looser orbits have mostly higher eccentricity, while in the simulations we find mixed outcomes.

This happens because the planet can escape before the star reaches its pericenter, thus invalidating assumption (i) of the analytic model. Moreover, $85 \%$ of the unbound planets begin the simulation outside $0.5 r_{\rm J}$ so that they may become immediately unbound and consequently violate all the assumptions of the analytic model.

The analytic model fails to predict the distribution of both semi-major axis and eccentricity in the case of retrograde orbits (set~B, Figure \ref{fig:ratios}, right-hand panel). This occurs because the escape mechanism for retrograde orbits is different from that of prograde orbits. Just a minor fraction of retrograde planets escape from one of the Lagrangian points (e.g. Figure~\ref{fig:aph}). Moreover, planets in retrograde orbits can survive several star pericenter passages before being kicked into an unstable orbit, and the planet escape may occur anywhere along the star orbit (see Figure \ref{fig:orbital}, right-hand panel).

Our results are consistent with the findings of \citet{sue13}, who studied the orbital properties of temporary captured planetesimals by a planet in circular heliocentric orbit. 
\citet{sue13} highlight that captures of planetesimals  into prograde orbits  about the planet (i.e., through L1 or L2 Lagrangian points) take place for a certain range of semi-major axes, leading to a gap in semi-major axis distribution, whereas captures into retrograde orbits do not produce a significant gap.
}


\begin{figure}[htbp]
	\begin{center}
		\includegraphics[width=1\columnwidth]{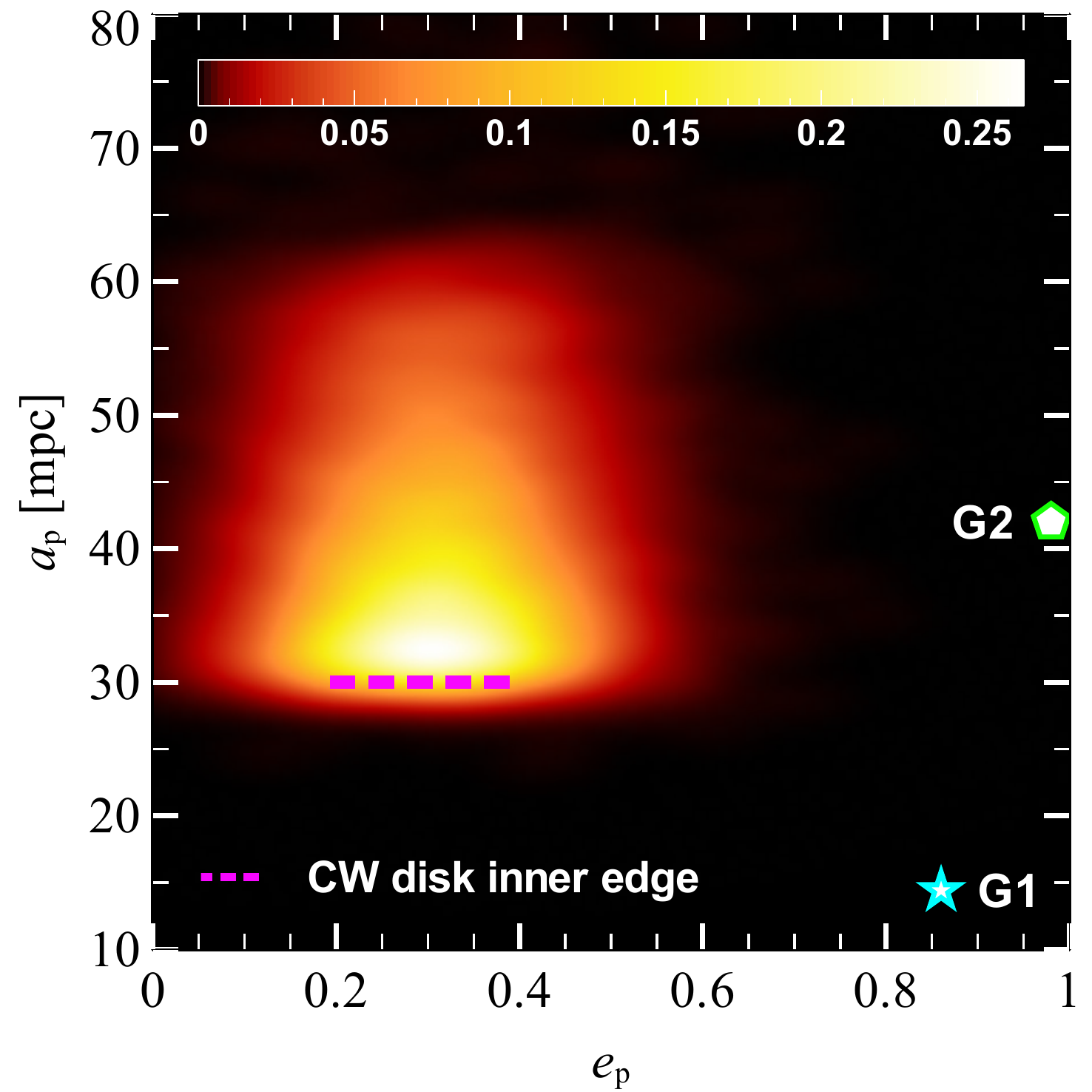}
		\caption{\label{fig:gorb} Probability density map of semi-major axis and eccentricity of captured planets in the CW disk simulations. Green pentagon: G2 cloud. Cyan star: G1 cloud. Magenta dashed line: inner edge of the CW disk. All simulated sets were used. %
		}
	\end{center}
\end{figure}

\begin{figure}[htbp]
	\begin{center}
		\includegraphics[width=1\columnwidth]{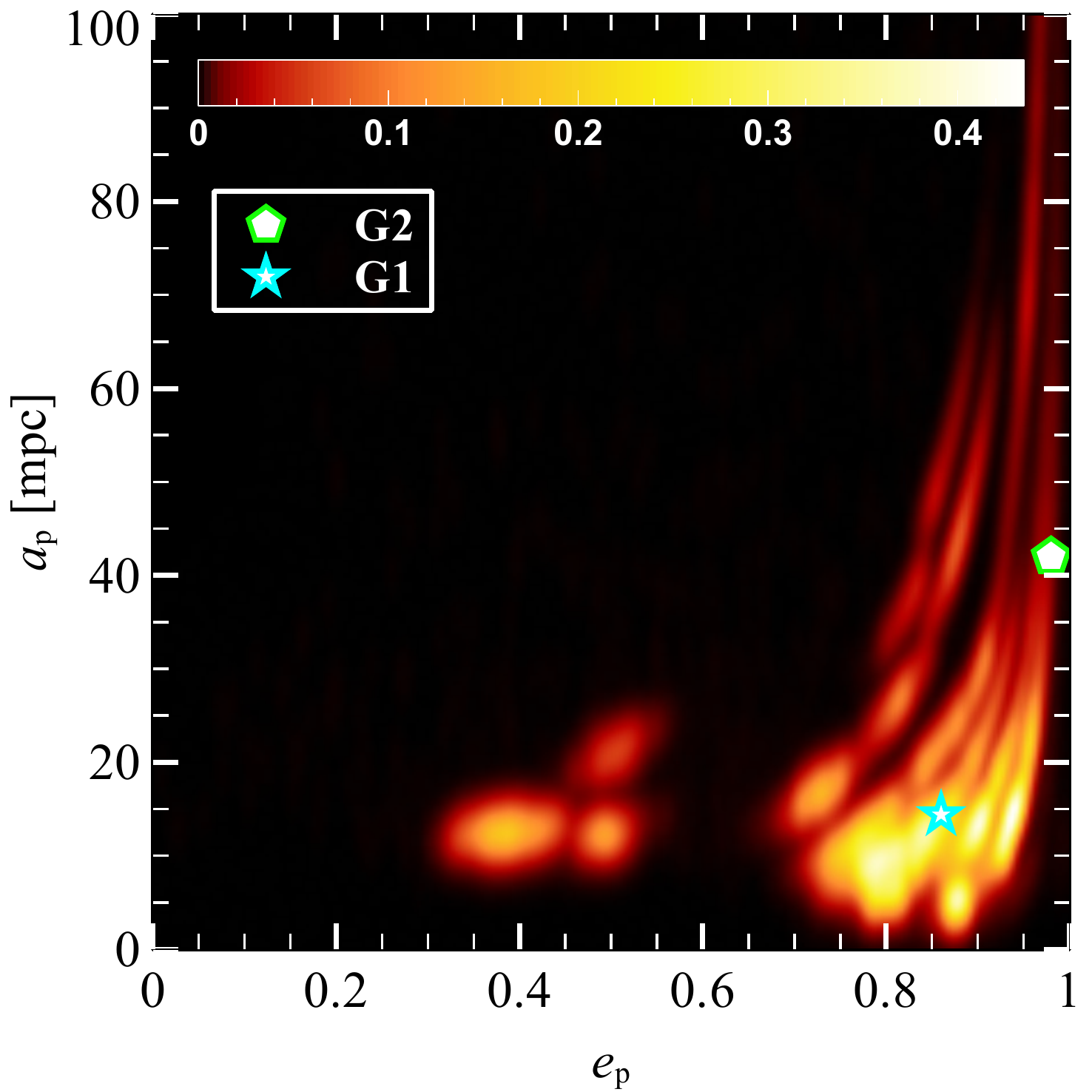}
		\caption{\label{fig:ssae} Probability density map of semi-major axis and eccentricity of the captured planets in the S-stars simulations. Green pentagon: G2 cloud. Cyan star: G1 cloud. Magenta dotted line: inner edge of the CW disk. %
		}
	\end{center}
\end{figure}

\begin{figure}[htbp]
	\begin{center}
		\includegraphics[width=1\columnwidth]{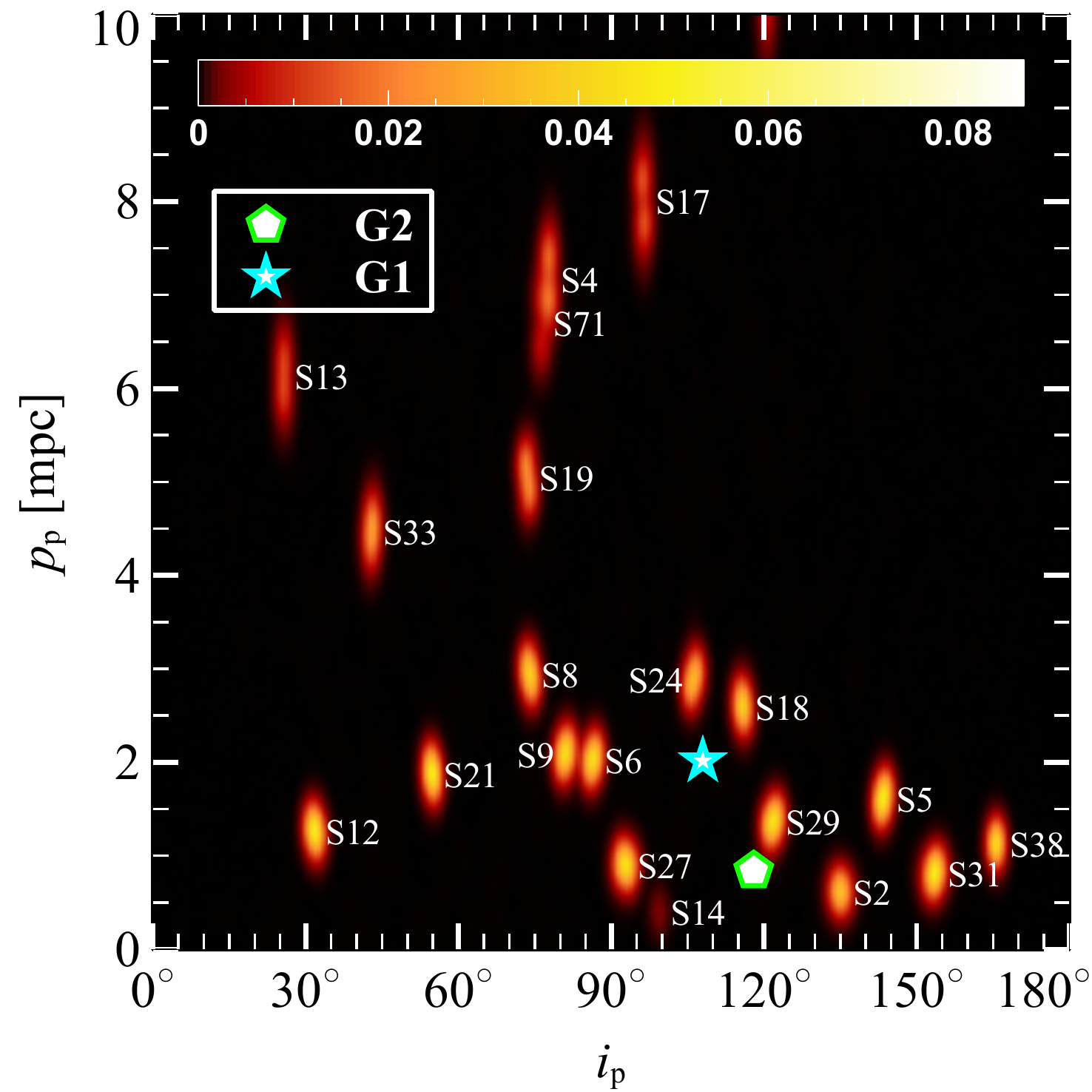}
		\caption{\label{fig:sspi} Probability density map of pericenter distance and inclination of captured planets in the S-stars simulations. Green pentagon: G2 cloud. Cyan star: G1 cloud. Each blob corresponds to the planets escaped by a single S-star, labeled on the map.%
		}
	\end{center}
\end{figure}

\subsection{Comparison with G2 and G1 cloud orbits}\label{sec:clouds}

Figure \ref{fig:gorb} shows the probability density map of finding an unbound planet in the semi-major axis -- eccentricity  plane for the CW disk simulations. No planet can match the orbits of the G1 or G2 cloud. In particular, none of the simulated planets can achieve a highly eccentric orbit.
In fact, the closest pericenter passage of an unbound planet in our simulations is $1750\rm\, AU$, a factor of $\sim 9$ larger than the pericenter passage of the G1 cloud.

Since unbound planets remain on orbits similar to those of their parent star, we expect that they will experience scattering with the stars in the CW disk. Angular momentum diffusion and scattering in the CW disk may bring low-mass objects on nearly radial orbits \citep{mur12}. N-body simulations that include the entire CW disk are required to study this effect and will be presented in a forthcoming study (Trani et al., in preparation).

Figure \ref{fig:ssae} is the same as Figure \ref{fig:gorb} but for captured planets in the S-stars simulations. Most planets escaped from the S-stars are on highly eccentric orbits and are compatible with the G1 and G2 cloud. In particular, we find that captured planets have a probability of $2\%$ and $70\%$ to have semi-major axis and eccentricity within $1\sigma$ of the observations for G2 and G1.

We also study the inclination of the orbits of captured planets.
Figure \ref{fig:sspi}  shows the probability density map of finding an unbound planet in the pericenter distance -- inclination plane for the S-star simulations. Since captured planets retain approximately the same inclination as their parent star, each blob corresponds to one or more S-stars.

None of the simulated planets has exactly the same inclination as G2 and G1 orbits. Although planets escaped from S29 lie very close to the position of G2 in the $p-i$ plane, further analysis reveals that longitude of the ascending node $\Omega$ mismatches by $\sim 75^\circ$; therefore the orbit of G2 and the one of the planets escaped from S29 do not lie on the same plane.
However,  the orbital properties of several S-stars are still unconstrained \citep{gil09a}. Many S-stars fainter than  $m_{\rm H}>19$ are not even detected. Identifying more S-stars and deriving their orbital properties (especially their inclinations) will give important clues on our scenario.


Moreover, explaining G1 and G2 with this scenario requires that planets can exist around S-stars. One of the most popular scenarios to explain the formation of the S-stars, the so-called binary breakup  scenario \citep{per09} predicts that the S-stars were captured by the SMBH via the Hills mechanism, during encounters with binary stars. 
A proto-planetary disk might be disrupted during the binary encounter with the SMBH. Alternatively, the planet might have been formed around the S-star before it was captured by the SMBH. \citet{gin12} showed that some planets will likely remain bound to their star during a three-body encounter, if their semi-major axis is $a^i_{\rm p} \gtrsim 0.5 \AU$, since planets with $a^i_{\rm p} \lesssim 0.5 \AU$ will be more likely ejected from the system. However, the closest the planet to the S-star, the more difficult is for the SMBH to capture it. All these issues deserve further study. 

Finally, we note that our simulations were done for a star-planet system, but our results can be generalized also to a star-star system. In other words, a low-mass star initially bound to an S-star might have been captured by the SMBH into a new orbit, matching the eccentricity and semi-major axis of G1 and G2.

\section{Conclusions}\label{sec:conclusions}

We investigated the dynamics of planets orbiting the young stars in the inner edge of the CW disk and in the S-star cluster by means of regularized N-body simulations.
We simulated $4\times{}10^4$ hierarchical systems consisting of the SMBH, a star and its planet lying in the CW disk.
We also ran $2\times{}10^4$ N-body realizations of the 27 innermost S-stars, assigning  a planet to each S-star.

The planet may escape its parent star and be tidally captured by the SMBH, depending on the properties of the orbit of the star and the planet. Planets on retrograde (prograde) orbits are captured if their orbit lies outside  $r_{\rm J}$ ($0.5 r_{\rm J}$), where $r_{\rm J}$ is the Jacobi radius.

We study the orbital properties of starless planets around the SMBH and find that planets remain on orbits similar to the ones of their parent star. In particular, we find that in $95 \%$ of the runs the semi-major axis and eccentricity of the planet orbit differ less than $6\%$ and $13\%$ from those of the parent star, respectively.

In case of prograde coplanar orbits, the semi-major axis of starless planets can be approximately predicted using a simple analytic model. We show that the escape mechanism of the planet from the  Hill sphere of the parent star determines the semi-major axis of the planet: if the planet escapes from the inner Lagrangian point (i.e. the one located towards the SMBH) it will end on a tighter orbit; in contrast, if the planet escapes from the outer Lagrange point it will end on a looser orbit. 
Furthermore, we find that looser orbits tend to have higher eccentricity with respect to the parent star orbit, while tighter orbits tend to have lower eccentricity.

In the case of planets in the CW disk, we find that the closest passage near the SMBH achieved by a starless planet is at $1750 \AU$, a factor $\sim 9$ larger than the pericenter distance of the G2 cloud orbit. We speculate that perturbations from other stars in the CW disk may bring planets into nearly radial orbits. In forthcoming studies we will investigate the effect of angular momentum transport and scatterings on the dynamics of planets in the CW disk.

In contrast, the semi-major axis and eccentricity of planets escaping from the S-stars  can match those of G1 and G2. The main issue is that the orbital planes of known S-stars do not match those of G1 and G2. Therefore, future detection of S-stars with approximately the same orbital plane as G1 and G2 are essential to support this scenario. We note that our simulations were run for star-planet systems, but our predictions apply to any low-mass companions of the CW disk stars and of the S-stars. Thus, our scenario also predicts that G1 and G2 might be low-mass stars that were previously bound to S-stars.

\acknowledgements
We thank the referee, Keiji Ohtsuki for his careful reading of the manuscript and for his invaluable comments that improved the manuscript.
The authors acknowledge financial support from INAF through grant PRIN-2014-14. MM and MS acknowledge financial support from the Italian Ministry of Education, University and Research (MIUR) through grant FIRB 2012 RBFR12PM1F. MM  acknowledges financial support from the MERAC Foundation. We thank Alessandro Ballone for useful and stimulating discussions.

\appendix
\section{Classification of planet orbits}\label{sec:app}
\begin{figure}[htbp]
	\begin{center}
		\includegraphics[width=0.49\linewidth]{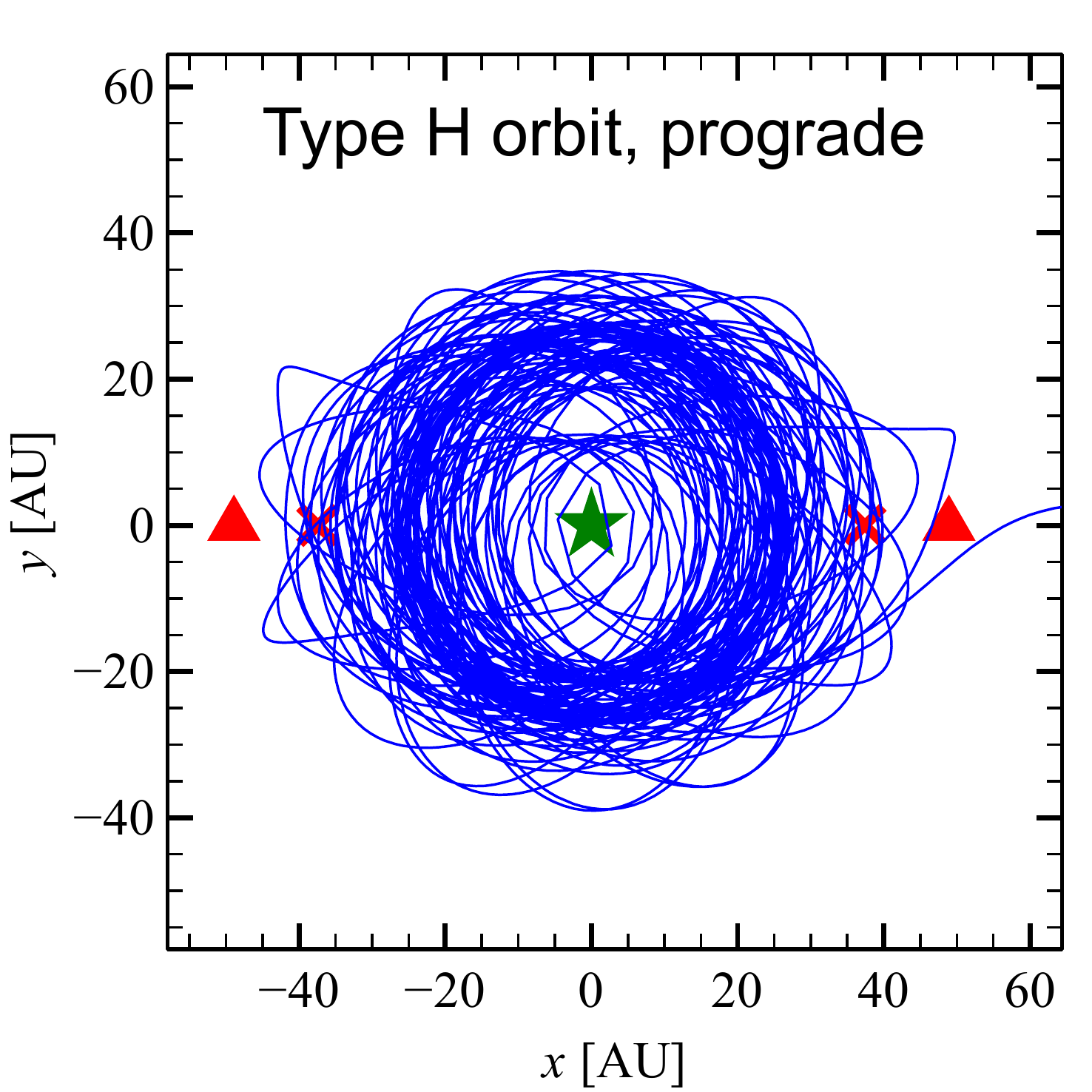}
		\includegraphics[width=0.49\linewidth]{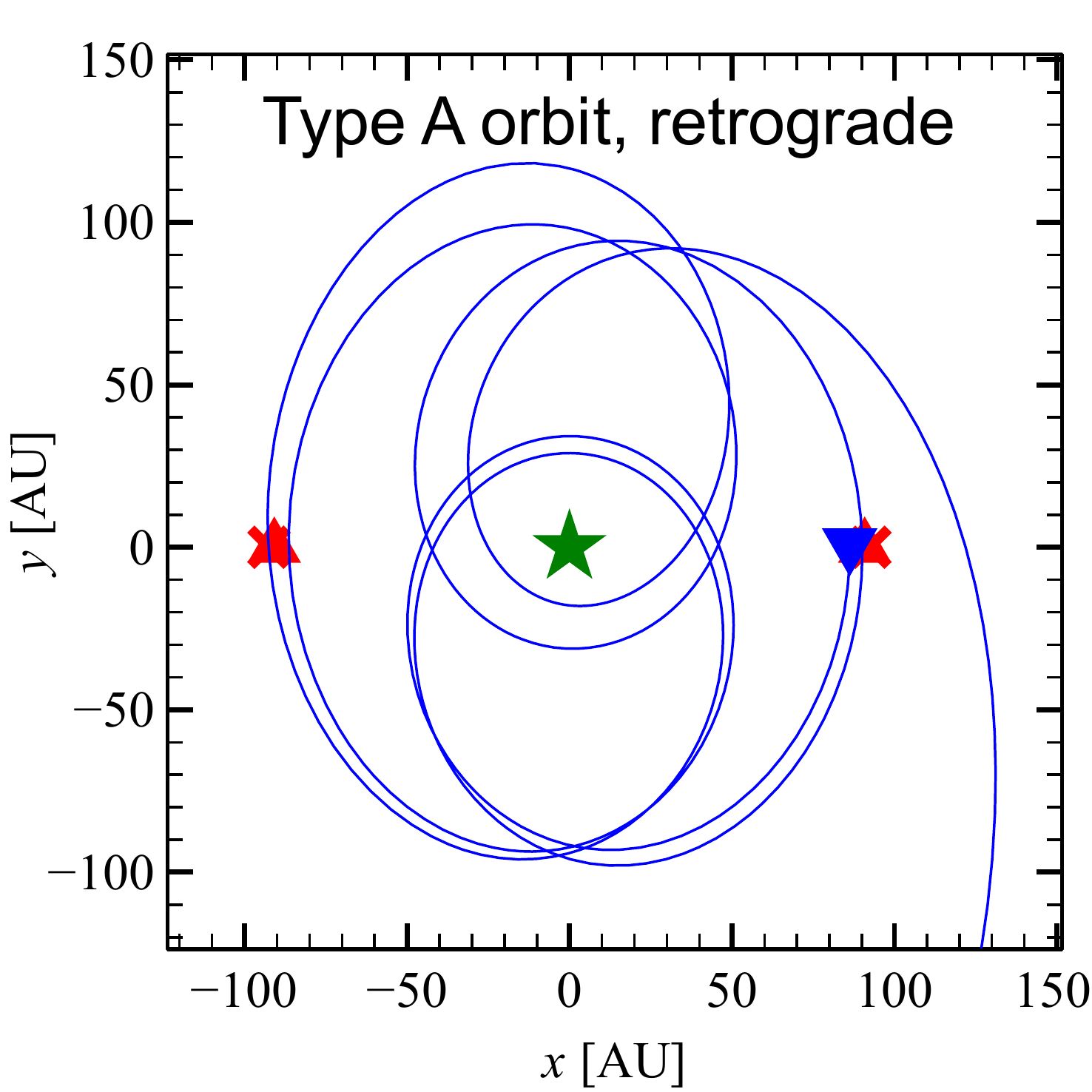}
		\caption{ \label{fig:class} Same as Figure~\ref{fig:orbital}, but for two different realizations. Left-hand panel: simulation from set~A (coplanar prograde) that we classify as Type~H \citep{sue11}. Right-hand panel: simulation from set~B (coplanar retrograde) that we classify as Type~A \citep{sue11}.
		}
	\end{center}
\end{figure}

{ 
Temporary planet orbits around the star (before tidal capture by the SMBH) can be classified according to \citet{sue11}, who studied the orbital properties of temporary captured planetesimals by a planet in circular heliocentric orbit. They distinguish four types of orbits, three for retrograde orbits and one for prograde orbits, and find that the orbit type depends on the eccentricity and energy of the planetesimal initial orbit around the Sun.
	
We find that most prograde orbits of set~A are of type~H (Hill sphere-shaped, left panel of Figure~\ref{fig:class}), which is typical of low-energy orbits that remain confined inside the Hill sphere, with escapes mainly occurring through the Lagrangian points.
On the other hand, most retrograde orbits of set~B are of type~A (apple-shaped, right panel of Figure~\ref{fig:class}). Planets on type~A orbits can orbit past the Hill sphere of the star without escaping. Escapes occur mainly in the SMBH-star direction but not strictly through the Lagrangian points.
	
We do not find any evidence of type~R~and~E orbits in our simulations. These orbit types were found by \citet{sue11} in the case of high velocity-dispersion between the planet and the planetesimal. The dispersion-dominated velocity regime is excluded by construction in our case, since the planet is initially bound to the star. 
	
We note that many orbits we examined are irregular and do not resemble any of aforementioned orbit types. This is due to the eccentricity of the star orbit that makes the tidal field experienced by the planet not stationary, unlike in the zero-eccentricity study of \citet{sue11}. This leads to an additional perturbation that can modify the shape of the planet orbit, and may cause earlier escape than in the zero-eccentricity case.}

\begin{thebibliography}{}
	\expandafter\ifx\csname natexlab\endcsname\relax\def\natexlab#1{#1}\fi

	\bibitem[{{Aarseth}(2003)}]{aarseth2003}
	{Aarseth}, S.~J. 2003, {Gravitational N-Body Simulations}, 430
	
	\bibitem[{{Alig} {et~al.}(2011){Alig}, {Burkert}, {Johansson}, \&
		{Schartmann}}]{alig2011}
	{Alig}, C., {Burkert}, A., {Johansson}, P.~H., \& {Schartmann}, M. 2011,
	\mnras, 412, 469
	
	\bibitem[{{Alig} {et~al.}(2013){Alig}, {Schartmann}, {Burkert}, \&
		{Dolag}}]{alig2013}
	{Alig}, C., {Schartmann}, M., {Burkert}, A., \& {Dolag}, K. 2013, \apj, 771,
	119
	
	\bibitem[{Ballone {et~al.}(2013)Ballone, Schartmann, Burkert, Gillessen,
		Genzel, Fritz, Eisenhauer, Pfuhl, \& Ott}]{bal13}
	Ballone, A., Schartmann, M., Burkert, A., {et~al.} 2013, {ApJ}, 776, 13
	
	\bibitem[{{Bartko} {et~al.}(2009){Bartko}, {Martins}, {Fritz}, {Genzel},
		{Levin}, {Perets}, {Paumard}, {Nayakshin}, {Gerhard}, {Alexander},
		{Dodds-Eden}, {Eisenhauer}, {Gillessen}, {Mascetti}, {Ott}, {Perrin},
		{Pfuhl}, {Reid}, {Rouan}, {Sternberg}, \& {Trippe}}]{bartko2009}
	{Bartko}, H., {Martins}, F., {Fritz}, T.~K., {et~al.} 2009, \apj, 697, 1741
	
	\bibitem[{{Bonnell} \& {Rice}(2008)}]{bonnell2008}
	  {Bonnell}, I.~A., \& {Rice}, W.~K.~M. 2008, Science, 321, 1060

	\bibitem[{{Cl\'enet et~al.}(2005)}]{cle05}
	  {Cl\'enet}, Y., Rouan, D., Gratadour, D., et al. 2005, A\&A, 439, L9

      \bibitem[{Burkert {et~al.}(2012)Burkert, Schartmann, Alig, Gillessen, Genzel,
		Fritz, \& Eisenhauer}]{bur12}
	Burkert, A., Schartmann, M., Alig, C., {et~al.} 2012, {ApJ}, 750, 58
	
	\bibitem[{{{\v C}ade{\v z}} {et~al.}(2008){{\v C}ade{\v z}}, {Calvani}, \&
		{Kosti{\'c}}}]{cadez2008}
	{{\v C}ade{\v z}}, A., {Calvani}, M., \& {Kosti{\'c}}, U. 2008, \aap, 487, 527
	
	\bibitem[{{Capuzzo-Dolcetta} {et~al.}(2013){Capuzzo-Dolcetta}, {Spera}, \&
		{Punzo}}]{capuzzo2013}
	{Capuzzo-Dolcetta}, R., {Spera}, M., \& {Punzo}, D. 2013, Journal of
	Computational Physics, 236, 580
	
	\bibitem[{Colle {et~al.}(2014)Colle, Raga, Contreras-Torres, \&
		Toledo-Roy}]{dec14}
	Colle, F.~D., Raga, A.~C., Contreras-Torres, F.~F., \& Toledo-Roy, J.~C. 2014,
	{ApJ}, 789, L33
	
	\bibitem[{Do {et~al.}(2013)Do, Lu, Ghez, Morris, Yelda, Martinez, Wright, \&
		Matthews}]{do13}
	Do, T., Lu, J.~R., Ghez, A.~M., {et~al.} 2013, {ApJ}, 764, 154
	
	\bibitem[{{Eckart} {et~al.}(2013){Eckart}, {Mu{\v z}i{\'c}}, {Yazici}, {Sabha},
		{Shahzamanian}, {Witzel}, {Moser}, {Garcia-Marin}, {Valencia-S.}, {Jalali},
		{Bremer}, {Straubmeier}, {Rauch}, {Buchholz}, {Kunneriath}, \&
		{Moultaka}}]{eck13}
	{Eckart}, A., {Mu{\v z}i{\'c}}, K., {Yazici}, S., {et~al.} 2013, \memsai, 84,
	618
	
	\bibitem[{Ghez {et~al.}(2003)Ghez, Duch{\^{e}}ne, Matthews, Hornstein, Tanner,
		Larkin, Morris, Becklin, Salim, Kremenek, Thompson, Soifer, Neugebauer, \&
		McLean}]{ghe03}
	Ghez, A.~M., Duch{\^{e}}ne, G., Matthews, K., {et~al.} 2003, \apj, 586, L127

	\bibitem[{Ghez {et~al.}(2005) Ghez, Hornstein, Lu, Bouchez, Le Mignant,  van Dam,  Wizinowich, Matthews, Morris, Becklin, Campbell, Chin, Hartman, Johansson, Lafon, Stomski, Summers}]{ghe05}
        Ghez, A.~M., Hornstein, S.~D., Lu, J.~R., {et~al.} 2005, \apj, 635, 1087
	
	\bibitem[{Gillessen {et~al.}(2009)Gillessen, Eisenhauer, Trippe, Alexander,
		Genzel, Martins, \& Ott}]{gil09a}
	Gillessen, S., Eisenhauer, F., Trippe, S., {et~al.} 2009, {ApJ}, 692, 1075
	
	\bibitem[{Gillessen {et~al.}(2011)Gillessen, Genzel, Fritz, Quataert, Alig,
		Burkert, Cuadra, Eisenhauer, Pfuhl, Dodds-Eden, Gammie, \& Ott}]{gil12}
	Gillessen, S., Genzel, R., Fritz, T.~K., {et~al.} 2011, Nature, 481, 51
	
	\bibitem[{{Gillessen} {et~al.}(2013a){Gillessen}, {Genzel}, {Fritz},
		{Eisenhauer}, {Pfuhl}, {Ott}, {Cuadra}, {Schartmann}, \& {Burkert}}]{gil13a}
	{Gillessen}, S., {Genzel}, R., {Fritz}, T.~K., {et~al.} 2013a, \apj, 763, 78
	
	\bibitem[{Gillessen {et~al.}(2013b)Gillessen, Genzel, Fritz, Eisenhauer, Pfuhl,
		Ott, Schartmann, Ballone, \& Burkert}]{gil13b}
	Gillessen, S., Genzel, R., Fritz, T.~K., {et~al.} 2013b, {ApJ}, 774, 44
	
	\bibitem[{Ginsburg {et~al.}(2012)Ginsburg, Loeb, \& Wegner}]{gin12}
	Ginsburg, I., Loeb, A., \& Wegner, G.~A. 2012, \mnras, 423, 948
	
	\bibitem[{Guillochon {et~al.}(2014)Guillochon, Loeb, MacLeod, \&
		Ramirez-Ruiz}]{gui14}
	Guillochon, J., Loeb, A., MacLeod, M., \& Ramirez-Ruiz, E. 2014, {ApJ}, 786, L12
        
	\bibitem[{{Hamers} \& {Portegies Zwart}(2015)}]{hamers2015}
	{Hamers}, A.~S., \& {Portegies Zwart}, S.~F. 2015, \mnras, 446, 710
	
	\bibitem[{Hamilton \& Burns(1991)}]{ham91}
	Hamilton, D.~P., \& Burns, J.~A. 1991, Icarus, 92, 118

	\bibitem[{Hamilton \& Burns(1992)}]{ham92}
	Hamilton, D.~P., \& Burns, J.~A. 1991, Icarus, 96, 43
	
	\bibitem[{{Hobbs} \& {Nayakshin}(2009)}]{hobbs2009}
	{Hobbs}, A., \& {Nayakshin}, S. 2009, \mnras, 394, 191
	
	\bibitem[{{Kosti{\'c}} {et~al.}(2009){Kosti{\'c}}, {{\v C}ade{\v z}},
		{Calvani}, \& {Gomboc}}]{kostic2009}
	{Kosti{\'c}}, U., {{\v C}ade{\v z}}, A., {Calvani}, M., \& {Gomboc}, A. 2009,
	\aap, 496, 307
	
	\bibitem[{{Lu} {et~al.}(2013){Lu}, {Do}, {Ghez}, {Morris}, {Yelda}, \&
		{Matthews}}]{lu2013}
	{Lu}, J.~R., {Do}, T., {Ghez}, A.~M., {et~al.} 2013, \apj, 764, 155
	
	\bibitem[{{Lu} {et~al.}(2009){Lu}, {Ghez}, {Hornstein}, {Morris}, {Becklin}, \&
		{Matthews}}]{lu2009}
	{Lu}, J.~R., {Ghez}, A.~M., {Hornstein}, S.~D., {et~al.} 2009, \apj, 690, 1463
	
	\bibitem[{{Lucas} {et~al.}(2013){Lucas}, {Bonnell}, {Davies}, \&
		{Rice}}]{lucas2013}
	{Lucas}, W.~E., {Bonnell}, I.~A., {Davies}, M.~B., \& {Rice}, W.~K.~M. 2013,
	\mnras, 433, 353
	
	\bibitem[{{Mapelli} \& {Gualandris}(2016)}]{mapelli2016b}
	{Mapelli}, M., \& {Gualandris}, A. 2016, in Lecture Notes in Physics, Berlin
	Springer Verlag, Vol. 905, Lecture Notes in Physics, Berlin Springer Verlag,
	ed. F.~{Haardt}, V.~{Gorini}, U.~{Moschella}, A.~{Treves}, \& M.~{Colpi}, 205
	
	\bibitem[{{Mapelli} {et~al.}(2008){Mapelli}, {Hayfield}, {Mayer}, \&
		{Wadsley}}]{mapelli2008}
	{Mapelli}, M., {Hayfield}, T., {Mayer}, L., \& {Wadsley}, J. 2008, ArXiv
	e-prints, arXiv:0805.0185
	
	\bibitem[{{Mapelli} {et~al.}(2012){Mapelli}, {Hayfield}, {Mayer}, \&
		{Wadsley}}]{mapelli2012}
	---. 2012, \apj, 749, 168
	
	\bibitem[{Mapelli \& Ripamonti(2015)}]{map15}
	Mapelli, M., \& Ripamonti, E. 2015, {ApJ}, 806, 197
	
	\bibitem[{{Mapelli} \& {Trani}(2016)}]{mapelli2016a}
	{Mapelli}, M., \& {Trani}, A.~A. 2016, \aap, 585, A161
	
	\bibitem[{Meyer \& Meyer-Hofmeister(2012)}]{mey12}
	Meyer, F., \& Meyer-Hofmeister, E. 2012, A\&A, 546, L2
	
	\bibitem[{{Mikkola} \& {Tanikawa}(1999{\natexlab{a}})}]{mikkola1999a}
	{Mikkola}, S., \& {Tanikawa}, K. 1999{\natexlab{a}}, \mnras, 310, 745
	
	\bibitem[{{Mikkola} \& {Tanikawa}(1999{\natexlab{b}})}]{mikkola1999b}
	---. 1999{\natexlab{b}}, Celestial Mechanics and Dynamical Astronomy, 74, 287
	
	\bibitem[{Miralda-Escud{\'{e}}(2012)}]{mir12}
	Miralda-Escud{\'{e}}, J. 2012, {ApJ}, 756, 86
	
	\bibitem[{Murray-Clay \& Loeb(2012)}]{mur12}
	Murray-Clay, R.~A., \& Loeb, A. 2012, Nature Communications, 3, 1049
	
	\bibitem[{{Paumard} {et~al.}(2006){Paumard}, {Genzel}, {Martins}, {Nayakshin},
		{Beloborodov}, {Levin}, {Trippe}, {Eisenhauer}, {Ott}, {Gillessen}, {Abuter},
		{Cuadra}, {Alexander}, \& {Sternberg}}]{paumard2006}
	{Paumard}, T., {Genzel}, R., {Martins}, F., {et~al.} 2006, \apj, 643, 1011
	
	\bibitem[{Pfuhl {et~al.}(2015)Pfuhl, Gillessen, Eisenhauer, Genzel, Plewa, Ott,
		Ballone, Schartmann, Burkert, Fritz, Sari, Steinberg, \& Madigan}]{pfu15}
	Pfuhl, O., Gillessen, S., Eisenhauer, F., {et~al.} 2015, {ApJ}, 798, 111

      \bibitem[{Perets {et~al.}(2009)Perets, Gualandris, Kupi, Merritt, \& Alexander}]{per09}
	Perets, H.~B., Gualandris, A., Kupi, G., Merritt, D., \& Alexander, T.  2009, \apj, 702, 884
        
	\bibitem[{Phifer {et~al.}(2013)Phifer, Do, Meyer, Ghez, Witzel, Yelda, Boehle,
		Lu, Morris, Becklin, \& Matthews}]{phi13}
	Phifer, K., Do, T., Meyer, L., {et~al.} 2013, {ApJ}, 773, L13
	
	\bibitem[{Prodan {et~al.}(2015)Prodan, Antonini, \& Perets}]{pro15}
	Prodan, S., Antonini, F., \& Perets, H.~B. 2015, {ApJ}, 799, 118
	
	\bibitem[{Schartmann {et~al.}(2012)Schartmann, Burkert, Alig, Gillessen,
		Genzel, Eisenhauer, \& Fritz}]{sch12}
	Schartmann, M., Burkert, A., Alig, C., {et~al.} 2012, {ApJ}, 755, 155
	
	\bibitem[{Schodel {et~al.}(2003)Schodel, Ott, Genzel, Eckart, Mouawad, \&
		Alexander}]{sch03}
	Schodel, R., Ott, T., Genzel, R., {et~al.} 2003, {ApJ}, 596, 1015
	
	\bibitem[{Scoville \& Burkert(2013)}]{sco13}
	Scoville, N., \& Burkert, A. 2013, {ApJ}, 768, 108
	
	\bibitem[{Shcherbakov(2014)}]{shc14}
	Shcherbakov, R.~V. 2014, {ApJ}, 783, 31
	
	\bibitem[{{Stoer} \& {Bulirsch}(2002)}]{stoer1980}
	{Stoer}, J., \& {Bulirsch}, R. 2002, {Introduction to Numerical Analysis}
	
	\bibitem[Suetsugu et~al.(2011)]{sue11}
	Suetsugu, R., Ohtsuki, K., Tanigawa, T. 2011, AJ, 142, 200

	\bibitem[Suetsugu et~al.(2013)]{sue13}
	Suetsugu, R., Ohtsuki, K. 2013, MNRAS 431, 1709
	
	\bibitem[{{Trani} {et~al.}(2016){Trani}, {Mapelli}, {Bressan}, {Pelupessy},
		{van Elteren}, \& {Portegies Zwart}}]{trani2016}
	{Trani}, A.~A., {Mapelli}, M., {Bressan}, A., {et~al.} 2016, \apj, 818, 29
	
	\bibitem[{{Williams} \& {Cieza}(2011)}]{williams2011}
	{Williams}, J.~P., \& {Cieza}, L.~A. 2011, \araa, 49, 67
	
	\bibitem[{Witzel {et~al.}(2014)Witzel, Ghez, Morris, Sitarski, Boehle, Naoz,
		Campbell, Becklin, Canalizo, Chappell, Do, Lu, Matthews, Meyer, Stockton,
		Wizinowich, \& Yelda}]{wit14}
	Witzel, G., Ghez, A.~M., Morris, M.~R., {et~al.} 2014, {ApJ}, 796, L8
	
	\bibitem[{{Yelda} {et~al.}(2014){Yelda}, {Ghez}, {Lu}, {Do}, {Meyer}, {Morris},
		\& {Matthews}}]{yelda2014}
	{Yelda}, S., {Ghez}, A.~M., {Lu}, J.~R., {et~al.} 2014, \apj, 783, 131
	
	\bibitem[{Yusef-Zadeh {et~al.}(2015)Yusef-Zadeh, Roberts, Wardle, Cotton,
		Sch\"{o}del, \& Royster}]{yus15}
	Yusef-Zadeh, F., Roberts, D.~A., Wardle, M., {et~al.} 2015, {ApJ}, 801, L26
	
	\bibitem[{{Zubovas} {et~al.}(2012){Zubovas}, {Nayakshin}, \&
		{Markoff}}]{zubovas2012}
	{Zubovas}, K., {Nayakshin}, S., \& {Markoff}, S. 2012, \mnras, 421, 1315
	
\end{thebibliography}

\end{document}